\newcommand{\ee}{\end{equation}}
\newcommand{\bea}{\begin{eqnarray}}
\newcommand{\eea}{\end{eqnarray}}
\newcommand{\bes}{\begin{subequations}}
\newcommand{\ees}{\end{subequations}}
\newcommand{\beg}{\begin{gather}}
\newcommand{\eeg}{\end{gather}}

\newcommand{\br}{{\bf r}}

\newcommand{\kp}{k^{\prime}}

\newcommand{\bei}{\begin{itemize}}
\newcommand{\ei}{\end{itemize}}
\newcommand{\bra}{\langle}
\newcommand{\ket}{\rangle}

\newcommand{\bsh}{\begin{shadebox}}
\newcommand{\esh}{\end{shadebox}}
\newcommand{\besh}{\mbox{ }\\ \begin{shadebox}}
\newcommand{\eesh}{\end{shadebox}\\ \mbox{ }}
\newcommand{\non}{\nonumber}

\newcommand{\bel}{\begin{list}{$\bullet$}{
\setlength{\itemsep}{0pt}
\setlength{\topsep}{5pt}}}
\newcommand{\el}{\end{list}}

\newcommand{\bec}{\begin{center}}
\newcommand{\eec}{\end{center}}
\newcommand{\ben}{\begin{enumerate}}
\newcommand{\een}{\end{enumerate}}

\newcommand{\eff}{e\!f\!f}

\def\dint12{\int \!\!d\br_1 \!\!\int \!\!d\br_2~}
\def\dpint12{\int \!\!d\br_1^{\prime} \!\!\int \!\!d\br_2^{\prime}~}

\makeatletter

\def\eqnarray{
  \stepcounter{equation}
  \let\@currentlabel=\theequation
  \global\@eqnswtrue
  \global\@eqcnt\z@
  \tabskip\@centering
  \let\\=\@eqncr
  $$\halign to \displaywidth\bgroup\@eqnsel\hskip\@centering
  $\displaystyle\tabskip\z@{##}$&\global\@eqcnt\@ne
  \hfill$\displaystyle{{}##{}}$\hfill
  &\global\@eqcnt\tw@$\displaystyle\tabskip\z@{##}$\hfill
  \tabskip\@centering&\llap{##}\tabskip\z@\cr}

\@addtoreset{equation}{section}
\renewcommand{\theequation}{\arabic{section}.\arabic{equation}}

\makeatother

\documentclass[byrevtex,pra,floatfix,nobibnotes,nofootinbib,10pt]{revtex4}

\usepackage{graphicx}
\usepackage{amsmath}

\makeatletter
\renewcommand{\theequation}{%
\thesection.\arabic{equation}} 
\@addtoreset{equation}{section} 
\makeatother

\begin{document}

\vspace*{3cm}

\title{
Rheological Study of Transient Networks with \\
Junctions of Limited Multiplicity
}

\author{Tsutomu~Indei}
\email{indei@fukui.kyoto-u.ac.jp}
\affiliation{
Fukui Institute for Fundamental Chemistry,
Kyoto University, Kyoto 606-8103, Japan \\
{\rm Tel:~+81-75-711-7907 \\
Fax:~+81-75-781-4757\\
}}


\maketitle


\bec
{\large Abstract}
\eec

\vspace*{.5cm}

Viscoelastic and thermodynamic properties of transient gels 
comprised of telechelic polymers are theoretically studied.
We extend classical theories of transient networks
so that correlations among polymer chains 
through the network junctions are taken into account.
This extension enables us to investigate
how rheological quantities, 
such as viscosity and elastic modulus,
are affected by the association equilibrium conditions,
and how these quantities are related to the aggregation number
of junctions.
We present a theoretical model of transient networks
with junctions comprised of variable number of hydrophobic
groups on the chain ends.
Elastically effective chains are 
defined as the chains whose both ends are associated with end groups
on other chains.
It is shown that the dynamic shear moduli 
are well described in terms of the Maxwell model
characterized by a single relaxation time and the high-frequency plateau modulus
as in the classical theories, but
the reduced dynamic shear moduli depend on 
the polymer concentration and temperature
through the reduced concentration $c$ given as a combination of
the association constant and the volume fraction of end groups.
The plateau modulus and the zero-shear viscosity
rise nonlinearly with increasing $c$ when $c$ is small,
but they are proportional to $c$ for higher $c$.
The relaxation time also increases as
$c$ increases due to the presence of pairwise junctions at small $c$.


\vspace*{3cm}

\bec

\vspace*{.5cm}

\noindent
thermoreversible gels /
associating polymers /
transient network theory /
junction-multiplicity/
linear rheology


\newpage


\section{Introduction}

In some polymer gels, 
junctions 
can break and recombine 
in thermal fluctuations or
under external forces.
They are called transient gels or physical gels.
Most transient gels
exhibit thermoreversible properties, i.e.,
they reversibly change state between gel and sol
as thermodynamic conditions vary.
Typically,
polymers forming
such transient thermoreversible gels 
carry
a small fraction of interacting groups
capable of forming bonds
due to associative forces such as hydrophobic interaction,
ionic association,
hydrogen bonding,
cross-linking by crystalline segments
and so on.
Among them,
hydrophobically-modified 
water-soluble amphiphilic 
polymers 
have attracted widespread interest in recent years \cite{winyek}.
Amphiphilic properties 
stem from
the hydrophilicity of the main chain and 
the hydrophobicity of the associative functional groups 
embedded in the main chain.
Attractive force among the functional groups induces the formation of
transient network in aqueous media above a certain
concentration.

One of the simplest class 
of associating polymers capable of forming a network
is the linear polymers having functional groups only at both ends.
They are called telechelic polymers.
Rheological properties
of these polymers
have been well studied from experimental 
\cite{jen1,annable1,jen2,tam,russel,ng,serero0,ma,ma2,serero,nicolai,mewis1,mewis2,kujawa} 
as well as theoretical 
\cite{tanaed1,tanaed2,wang,mar0,mar1,semenov,meng,innon}
point of view with an intention of obtaining a
fundamental understanding of associating polymer systems.
Examples of telechelic polymers are
poly(ethylene oxide) (PEO) chains end-capped with short alkyl groups
\cite{jen1,annable1,jen2,winnik,winnik1,tam,russel0,russel,ng,ma,ma2,nicolai,mewis1,meng},
perfluoroalkyl end-capped PEO
\cite{serero0,serero}
and telechelic poly({\it N}-isopropylacrylamide) (PNIPAM) 
carrying octadecyl groups at both ends \cite{kujawa}.
They
exhibit characteristic rheological properties such as
temperature-frequency superposition onto a Maxwell fluid 
\cite{annable1}, 
breakdown of the Cox-Merz rule \cite{annable1, mewis1}, 
strain hardening \cite{serero0,mewis1},
shear thickening at relatively low shear rate followed by 
shear thinning \cite{jen1,annable1,serero0,tam,ma}, 
etc \cite{mewis1,mewis2}. 

In order to investigate molecular origin
of these phenomena,
Tanaka and Edwards (referred to as TE in the following) developed
the theory for transient networks \cite{tanaed1,tanaed2}
by extending the kinetic theory for reacting polymers \cite{gt}.
Under the Gaussian chain assumption,
TE succeeded to explain,
for example,
the linear response to the small oscillatory deformation
described in terms of the Maxwell model with a single relaxation time.
Shear thickening can be also explained
by extending the TE theory so that
the tension along the middle chain
contains a nonlinear term
\cite{innon}.
We can also 
treat
trifunctional associating polymers carrying two different species
of functional groups 
by an straightforward extension of the TE theory \cite{intana1,intana2} .
Several theories to 
treat
dynamic properties of transient networks
have also been presented up to now. 
For example, Wang \cite{wang} 
took isolated chains into consideration,
and Vaccaro and Marrucci \cite{mar1} 
incorporated the effect of incomplete relaxation of detached chains.

\begin{figure}[t]
\begin{center}
\includegraphics*[scale=0.5]{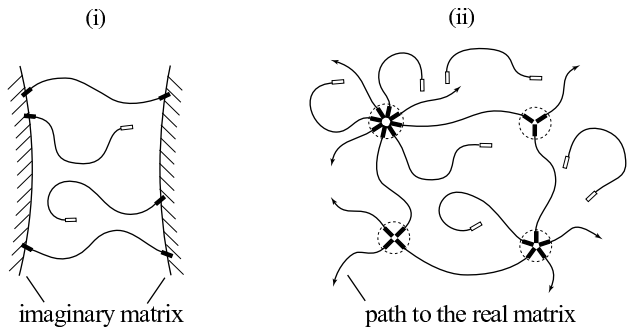}
\end{center}
\vspace*{-0.5cm}
\caption{(i) Drawing of the "network"
postulated in the conventional transient network theories.
Each chain interacts 
with fictitious matrix.
(ii) Schematic of the network supposed in the present series
of papers. 
Chains interact with each other through 
dissociation/association among end groups.
Arrows indicate paths to the (real) network matrix.
}
\label{fig0}
\end{figure}

In all transient network theories proposed so far
\cite{tanaed1,tanaed2,wang,mar0,mar1,intana1,intana2,innon},
it is implicitly assumed that a fictitious network 
exists a priori (see Fig.\ref{fig0} (i)).
This network matrix is not a substantial one in a sense that 
it itself does not contribute to the elasticity of the system, 
but it plays a role as a substrate of the chains 
on which association/dissociation of the end groups takes place.
Chains whose both ends are connected with this matrix
are supposed to be elastically effective.
Correlations among chains are not taken into account 
in this treatment because each chain interacts only with such matrix, and 
consequently
the concentration dependence of the rheological quantities
cannot be properly predicted; 
the elastic modulus, viscosity etc are 
simply proportional to the polymer concentration.
Furthermore, it is difficult by definition to incorporate the 
information about the network junction such as the aggregation number
and to study the effect of surfactants added to the system.

In this series of papers, we remove
this unfavorable assumption 
and develop a theory of thermoreversible transient networks formed by 
multiple junctions comprised of 
any number of functional groups
as depicted in Fig.\ref{fig0} (ii).
This modification enable us to 
deal with the sol/gel transition phenomenon and to
predict the proper concentration dependence of the dynamic shear moduli 
$G^{\prime}$ and $G^{\prime\prime}$ in the postgel regime 
of the solution.
In the following,
we refer to the number of functional groups per a junction 
(so called aggregation number)
as the junction multiplicity according to ref. \cite{tanastock}.

In this paper,
we formulate a theory of transient network
formed by thermoreversible multiple junction.
Elastically effective chains, or active chains, 
are defined only in local,
i.e., chains whose both ends are 
associated with other functional groups
are supposed to be elastically effective.
We derive a formula to calculate $G^{\prime}$ and $G^{\prime\prime}$
as a function of the frequency $\omega$ of applied small oscillation,
and analyze how 
$G^{\prime}$ and $G^{\prime\prime}$ characterized by
the high-frequency plateau modulus $G_{\infty}$, 
zero-shear viscosity $\eta_0$
and relaxation time $\tau$ depend on the polymer concentration.
It is shown that $G_{\infty}$ and $\eta_0$ increase nonlinearly with concentration
at lower concentration
and $\tau$ also increases with concentration 
corresponding to
experimental observation \cite{annable1,russel,meng}.
In equilibrium, 
we find the following mass action law:
$\psi_k=K_k(\psi_1)^k$ 
under the assumption that 
the connection rate of a functional group 
to a junction with the multiplicity $k$ (called $k$-junction) is
proportional to the number of 
functional groups in
$k$-junctions.
($\psi_k$ is the volume fraction of the functional groups belonging to $k$-junctions,
and $K_k$ is the reaction constant for the formation of a $k$-junction
from $k$ isolated functional groups.)
This relation is equivalent to the multiple-equilibrium condition that 
Tanaka and Stockmayer derived
in the theory of associating polymer solutions \cite{tanastock}.
In order to investigate the critical behavior of 
$G^{\prime}$ and $G^{\prime\prime}$
near the gelation point,
we need to define active chains taking account of global structure of the network.
Studies on the basis of this definition for active chains
will be presented in the 
following paper.
Looped chains, and then flower micelles comprised of loops, 
are assumed to be absent for simplicity
throughout this series.
Effects of the competition between the intra- and intermolecular association
at a junction 
as well as the effect of 
additives such as surfactants \cite{fransur,annable2,winniksur} 
will be investigated in the forthcoming papers.


This paper is organized as follows.
In section \ref{secII}, we will derive time development equations for
the distribution function of 
chains whose one end is incorporated into a $k$-junction while the other end
is belonging to a $\kp$-junction.
Kinetic equations for these chains will be also derived in this section.
In section \ref{secrates}, 
we will introduce the dissociation/association rates of functional groups.
Equilibrium 
distribution of the junction multiplicity and its mean value 
as a function of polymer concentration
will be discussed in section \ref{eqiv}.
In section \ref{secv}, we will study linear rheology of the system.
Finally, a relation between the present theory and the TE theory will be 
discussed in section \ref{secvidis}.


\section{Time Development of Transient Networks formed by 
Junctions with Variable Multiplicity}
\label{secII}

\subsection{Assumptions}

We consider athermal solutions of linear polymers (or primary chains)
carrying two functional groups at their both ends.
Here, functional group is a groups or a short segment of the primary chain
that can form aggregates (or junctions) 
in a solution through the noncovalent bonding. 
Primary chains can associate with each other 
in a solution through the aggregation of functional groups,
while they can be detached from others due to 
thermal agitation or macroscopic deformations applied to the system.
In equilibrium,
thermodynamic conditions (temperature, polymer concentration etc)
determine the association/dissociation 
rate of functional groups and hence the amount 
of junctions in the solution.
In this sense, these junctions are thermoreversible.
Above a certain polymer concentration,
primary chains eventually construct a 
macroscopic network
physically cross-linked by these junctions.
Thus the network we are going to study is also
thermoreversible as well as transient.
The association-dissociation reaction 
of functional groups is assumed to occur step by step.

We allow junctions to be formed by any number of functional groups.
According to Tanaka and Stockmayer \cite{tanastock},
the number of functional groups forming a junction 
is referred to as the junction multiplicity.
We also call the junction of the multiplicity $k$ $(=\!1,2,3,\cdots)$
the $k$-junction, i.e., 
$1$-junction is an unreacted group (or a functional group 
which remains unassociated), $2$-junction is a pairwise junction, etc.
For a while, let us
identify a head and a tail of each chain for convenience
by marking one end of each chain temporarily.
Of course, this 
does not affect mechanical properties of our system.
Then we term the primary chain whose head is 
incorporated into a $k$-junction whereas tail is a member of the 
distinct $\kp$-junction as the $(k,\kp)$-chain.
For instance, $(k,1)$-chain ($k\!\ge\!2$) is
a (primary) dangling chain whose tail is not connected with 
any other groups, and $(1,1)$-chain is an isolated chain.

We assume that chains are elastically effective when their both ends are 
connected with any other functional groups.
These chains are called active chains.
Active chains are assumed to
deform affinely to the macroscopic deformation applied to the system.
Note that active chains are defined only in local 
in a sense that chains are 
elastically effective
irrespective of whether 
functional groups they are connected with
belong to the infinite network or not.
As shown below, we can study
the sol/gel transition phenomenon
only in unsatisfactory in this theoretical framework.
The definition of active chains will be extended 
in the following paper
so that global information of the network can be included.
The Rouse relaxation time $\tau_R$ of the primary chain
is assumed to be much smaller than 
the characteristic time of a macroscopic deformation
applied to the system
and the lifetime of active chains,
so that chains in elastically non-effective (i.e., dangling and isolated) states 
are virtually in an equilibrium state even under flow.
The primary chains are assumed to be Gaussian 
with uniform molecular weight $M$ (number of repeat units is $N$)
that is smaller than the entanglement molecular weight.


\subsection{Time-Development Equation for Active Chains}

Let $F_{k,\kp}(\br,t)d\br$ be the number of $(k,\kp)$-chains 
at time $t$
per unit volume having the head-to-tail vector $\br\sim\br+d\br$.
Then the total number $\nu_{k,\kp}(t)$ of $(k,\kp)$-chains 
(per unit volume) is given by 
\bea
\nu_{k,\kp}(t)=\int\!d\br~ F_{k,\kp}(\br,t).
\eea
Dangling 
and isolated chains 
are the Gaussian chain irrespective of time, that is,
\begin{subequations}
\bea
\hspace{2.2cm}
& &F_{1,\kp}(\br,t)=\nu_{1,\kp}(t)f_0(r) ~~~~~(\mbox{for}~k\ge 1), \\
& &F_{k,1}(\br,t)=\nu_{k,1}(t)f_0(r) ~~~~~(\mbox{for}~\kp\ge 1),
\eea
\end{subequations}
where
\bea
f_0(\br)\equiv\left(\frac{3}{2\pi Na^2}\right)^{3/2}
\exp\left(-\frac{3|\br|^2}{2Na^2}\right)
\eea
is the probability distribution function (PDF) that
the chain takes the end-to-end vector $\br$
($a$ is the length of a repeat unit of the chain).
The number of chains whose head is incorporated into a $k$-junction is given by
\bea
\chi^{(h)}_{k}(t)=\sum_{l\ge1}\nu_{k,l}(t),
\eea
while the number of chains whose tail is belonging to a $k$-junction is 
\bea
\chi^{(t)}_{k}(t)=\sum_{l\ge1}\nu_{l,k}(t).
\eea
Then the number of chains whose one end, irrespective of whether it is 
a head or tail,
is incorporated into a $k$-junction (called $k$-chain hereafter) 
can be expressed as
\bea
\chi_{k}(t)=\frac{1}{2}\left(\chi^{(h)}_{k}(t)+\chi^{(t)}_{k}(t) \right).
\label{defwateigi}
\eea
A factor $1/2$ is necessary to avoid double counting of the $k$-chain.
Our aim in this section is to find time-development equations that 
$F_{k,\kp}(\br,t)$, $\nu_{k,\kp}(t)$ and $\chi_{k}^{(\cdot)}(t)$ 
(dot represents $h$ or $t$) 
obey when a macroscopic deformation
represented by the rate of deformation tensor 
$\hat{\kappa}(t)$ is applied to the system.

Firstly, we derive the equation for active 
$(k,\kp)$-chains ($k,\kp \ge 2$).
According to the affine deformation assumption,
$F_{k,\kp}(\br,t)$
satisfies the following evolution equation:
\bea
\frac{\partial F_{k,\kp}(\br,t)}{\partial t}
+\nabla\cdot\bigl(\dot{\br}(t)F_{k,\kp}(\br,t)\bigr)=
W^{(h)}_{k,\kp}(\br,t)+W^{(t)}_{k,\kp}(\br,t)~~~(\mbox{for}~k,\kp\ge 2),
\label{activeeq}
\eea
where $\dot{\br}(t)=\hat{\kappa}(t)\br$
is the rate of deformation of the 
head-to-tail vector $\br$ of the active chain.
The first (or the second) term in the right-hand side of (\ref{activeeq})
is a reaction term that describes the net increment of $F_{k,\kp}(\br,t)$
per unit time due to association-dissociation reactions among 
the head (or tail) of the $(k,\kp)$-chain and the functional groups
on the other chains.
\begin{figure}[t]
\begin{center}
\includegraphics*[scale=0.6]{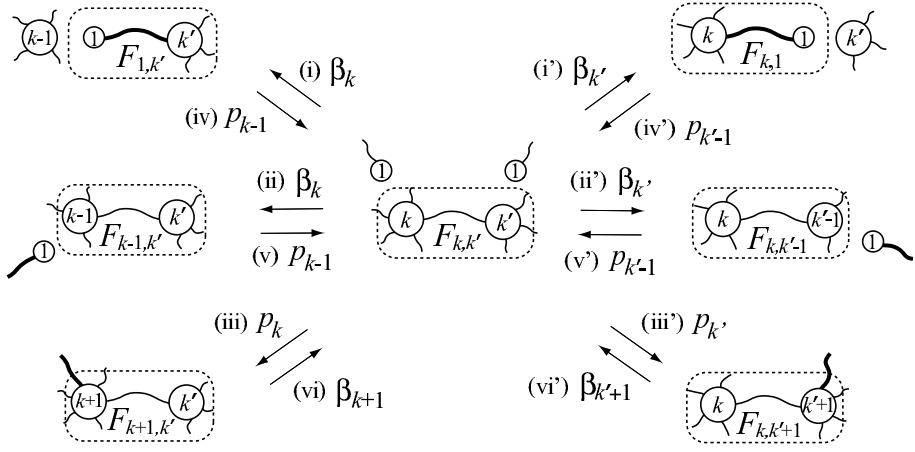}
\end{center}
\vspace*{-0.5cm}
\caption{
Reaction between the functional group on the $(k,\kp)$-chain and 
the functional group on the other chains.
Circles indicate junctions and lines 
emanating from circles represent the primary chains.
The letter inside the circle denotes the junction multiplicity.
Smaller circles with the number 1 inside represent unreacted functional groups.
Chains depicted by bold lines participate in the reaction.
The number of $(k,\kp)$-chains change if one of these (stepwise) reaction occur.
Reaction rates are shown near arrows for each reaction.
}
\label{tranew}
\end{figure}
The kinetic equation for active $(k,\kp)$-chains is then given by
integrating both sides of (\ref{activeeq}) with respect to $\br$.
We find
\bea
\frac{d\nu_{k,\kp}(t)}{dt}=w^{(h)}_{k,\kp}(t)+w^{(t)}_{k,\kp}(t),
\label{kineticeq}
\eea
where 
\bea
w^{(\cdot)}_{k,\kp}(t)\equiv\int d\br~W^{(\cdot)}_{k,\kp}(\br,t)
~~~~(\mbox{for}~k,\kp\ge 2).
\eea
%
%
The reaction terms $W^{(h)}_{k,\kp}(\br,t)$ and $w^{(h)}_{k,\kp}(t)$
are derived according to the following procedure.
The number of active $(k,\kp)$-chains 
decreases if\footnote{We are considering reactions 
regarding the head of the $(k,\kp)$-chain. 
The multiplicity of the junction to which its tail is connected is
assumed to be fixed at $\kp$.
} 
\begin{list}{}{}
\renewcommand{\makelabel}{}
\item[~~(i)]
the head of the $(k,\kp)$-chain is dissociated from a $k$-junction, or
\item[~~(ii)]
a functional group on the other chain
is dissociated from the head of the $(k,\kp)$-chain, or
\item[~~(iii)]
an unreacted functional group connects with the head of the $(k,\kp)$-chain. 
\end{list}
On the other hand, the number of active $(k,\kp)$-chains increases if 
\begin{list}{}{}
\renewcommand{\makelabel}{}
\item[~~(iv)]
the unreacted head of the $(1,\kp)$-chain is connected with 
$k\!-\!1$-junction, or
\item[~~(v)]
the unreacted group of the other chain
is connected with a head of the $(k\!-\!1,\kp)$-chain, or
\item[~~(vi)]
a functional group on the other chain is disconnected from the head 
of the $(k\!+\!1,\kp)$-chain. 
\end{list}
These reactions (i) $\sim$ (vi) are schematically depicted in Fig.\ref{tranew}
where the corresponding reactions regarding the tail of the $(k,\kp)$-chain,
(i') $\sim$ (vi'), are also shown.
\begin{figure}[t]
\begin{center}
\includegraphics*[scale=0.6]{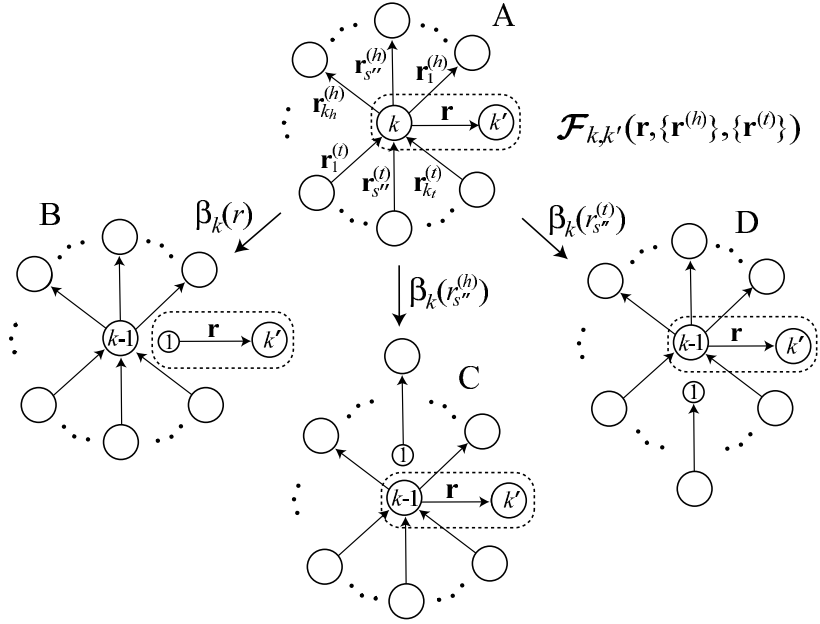}
\end{center}
\vspace*{-0.5cm}
\caption{
Explanation of eq. (\ref{statenumber}).
Circles stand for junctions.
Arrows connecting two junctions 
represent the head-to-tail vector of the chain
(chains are not shown for simplicity).
The number 
of the state A is given by (\ref{statenumber2}).
The first term in the parenthesis of (\ref{statenumber})
describes
the transition from the state A to the state B,
while 
the second (or the third) term in the parenthesis of (\ref{statenumber})
represents the transition from A to C (or to D).
}
\label{hitotsu}
\end{figure}
%
%
%
Now we introduce the probability (or dissociation rate) 
$\beta_k(r)$ that a functional group 
incorporated into the $k$-junction ($k\ge 2$) detaches itself from the junction 
per unit time.
In general, it is an increasing function with respect to the end-to-end length 
$r=|\br|$ of the $k$-chain \cite{tanaed1}.
Then the increment of $(k,\kp)$-chains with the head-to-tail vector $\br$ 
due to the reactions (i) and (ii) per unit time is written as
\bea
\int \prod_{s=1}^{k_h}d\br_s^{(h)}
\int \prod_{s^{\prime}=1}^{k_t}d\br_s^{(t)}
\left(
\beta_k(r)+
\sum_{s^{\prime\prime}=1}^{k_h} \beta_k\bigl(r^{(h)}_{s^{\prime\prime}}\bigr)
+\sum_{s^{\prime\prime}=1}^{k_t} \beta_k\bigl(r^{(t)}_{s^{\prime\prime}}\bigr)
\right)
{\cal F}_{k,\kp}\bigl(\br,\{\br^{(h)}\},\{\br^{(t)}\};t\bigr),
~~~~~~~~
\label{statenumber}
\eea
where the first term $\beta_k(r)$ in the parenthesis stems from the reaction (i)
while the second and the third terms are originated from the reaction (ii).
${\cal F}_{k,\kp}\bigl(\br,\{\br^{(h)}\},\{\br^{(t)}\};t\bigr)$
is the number of 
$(k,\kp)$-chains with the head-to-tail vector $\br$
whose head is incorporated into the $k$-junction formed by
$k_h$ heads of the other chains each having the head-to-vector 
$\br_1^{(h)},\cdots,\br_{k_h}^{(h)}$
and $k_t$ tails of the other chains 
each having the head-to-vector 
$\br_1^{(t)},\cdots,\br_{k_t}^{(t)}$
(see Fig.\ref{hitotsu}).
Note that a relation $(k_h+1)+k_t=k$ holds.
We can express ${\cal F}_{k,\kp}$ as follows:
\bea
{\cal F}_{k,\kp}(\br~,\{\br_s^{(h)}\},\{\br_{s^{\prime}}^{(t)}\};t)=
C_{k_h}
F_{k,\kp}(\br,t)
\prod_{{s}=1}^{k_h}f_k^{(h)}(\br_{s}^{(h)},t)
\prod_{{s^{\prime}}=1}^{k_t}f_k^{(t)}(\br_{s^{\prime}}^{(t)},t),
\label{statenumber2}
\eea
where
\bea
f_{k}^{(h)}(\br,t)\equiv
\frac{\sum_{l\ge1} F_{k,l}(\br,t)}{\sum_{l\ge1} \nu_{k,l}(t)}
\label{pdfdefh}
\eea
is the PDF that 
the chain whose head is incorporated into a $k$-junction
has the head-to-tail vector $\br$, and similarly
\bea
f_{k}^{(t)}(\br,t)\equiv
\frac{\sum_{l\ge1} F_{l,k}(\br,t)}{\sum_{l\ge1} \nu_{l,k}(t)}
\label{pdfdeft}
\eea
is the PDF that the chain whose tail is belonging to a $k$-junction
takes the head-to-tail vector $\br$.
Note that $f_{k}^{(\cdot)}(\br,t)$ 
is normalized to one, i.e.,
\bea
\int \!d\br ~f_{k}^{(\cdot)}(\br,t)=1.
\eea
A prefactor $C_{k_h}$ is the probability that the $k$-junction is formed by
$k_h$ heads and $k_t$ tails (in addition to a head of the ($k,\kp$)-chain),
and hence it is given by
\bea
C_{k_h}=\frac{1}{2^{k-1}}\frac{(k-1)!}{k_h!k_t!}.
\label{probkh}
\eea
Upon substitution of (\ref{statenumber2}) 
together with
(\ref{pdfdefh}), (\ref{pdfdeft}) and (\ref{probkh}),
eq. (\ref{statenumber}) reduces to
\bea
\beta_k(r)F_{k,\kp}(\br,t)+
(k-1)
\bra \beta_k(r) \ket(t)
F_{k,\kp}(\br,t),
\label{herusiki00}
\eea
where 
\bea
\bra \beta_k(r) \ket(t)\equiv  \int d\br~\beta_k(r)f_k(\br,t)
\label{betakitai}
\eea
is the expectation value of $\beta_k(r)$ averaged 
with respect to the PDF for $k$-chains given by
\bea
f_k(\br,t)\equiv\frac{1}{2}\left(f_k^{(h)}(\br,t)+f_k^{(t)}(\br,t)\right).
\eea
%
%
The number of $(k,\kp)$-chains decreases through the reaction (iii) 
only when the $k$-junction, to which an unreacted group are going to connect,
includes the head of a $(k,\kp)$-chain (with head-to-vector $\br$).
Here let us introduce the probability (or connection rate) $p_k(t)$ 
that an unreacted functional group 
catches a $k$-junction per unit time.\footnote{In general, 
$p_k(t)$ depends on the number $\mu_k(t)$ of the $k$-junction in the system;
the more the $k$-junction, the higher the probability for an unreacted group to catch 
the junction (see section \ref{secrates} for detail).
Thus $p_k(t)$ depends on time through $\mu_k(t)$.
}
Then the decrement of $F_{k,\kp}(\br,t)$ due to the reaction (iii)
is written as the product of the number $p_k(\chi_1^{(h)}(t)+\chi_1^{(t)}(t))$
of unreacted ends (both head and tail) that catches the $k$-junction per unit time 
and the number of $(k,\kp)$-chain with head-to-vector $\br$ 
per a $k$-junction, that is,
\bea
p_k(t)\left(\chi_{1}^{(h)}(t)+\chi_{1}^{(t)}(t)\right)
\frac{F_{k,\kp}(\br,t)}{\mu_k(t)}
=kp_k(t)\frac{\chi_1(t)}{\chi_k(t)}F_{k,\kp}(\br,t),
\label{huerusiki}
\eea
where 
\bea
\mu_k(t)\equiv\frac{\chi_k^{(h)}(t)+\chi_k^{(t)}(t)}{k}.
\label{muteigids}
\eea
is the number of $k$-junctions 
and a relation (\ref{defwateigi}) has been used for the equality.
%
%
The increment of the $(k,\kp)$-chain with the head-to-tail vector $\br$
through the reaction (iv) is given by
\bea
p_{k-1}(t)F_{1,\kp}(\br,t).
\eea
The increment of $F_{k,\kp}(\br,t)$ due to the reaction (v) 
is obtained by replacing $k$ in (\ref{huerusiki}) with $k-1$
as follows:
\bea
(k-1)p_{k-1}(t)\frac{\chi_1(t)}{\chi_{k-1}(t)}F_{k-1,\kp}(\br,t),
\label{huerusiki2}
\eea
whereas
the increment due to the reaction (vi) is given by
replacing $k$ in the second term of (\ref{herusiki00}) with $k+1$, i.e.,
\bea
k\bra \beta_{k+1}(r)\ket(t)
F_{k+1,\kp}(\br,t).
\label{herusiki1}
\eea
Summarizing, the reaction term regarding the head of the ($k,\kp$)-chains 
found to be
\bea
W^{(h)}_{k,\kp}(\br,t)=
& &-\beta_k(r)F_{k,\kp}(\br,t)+p_{k-1}(t)F_{1,\kp}(\br,t) \non\\
& &-B_k(t)F_{k,\kp}(\br,t)+B_{k+1}(t)F_{k+1,\kp}(\br,t) \non\\
& &-P_k(t)F_{k,\kp}(\br,t)+P_{k-1}(t)F_{k-1,\kp}(\br,t)
~~~~(\mbox{for}~k,\kp\ge 2)
\label{hannouactive}
\eea
where we have put
\begin{subequations}\label{BPteigi}
\bea
& &B_k(t)\equiv (k-1)
\bra \beta_{k}(r)\ket(t), 
\\
& &P_k(t)\equiv kp_k(t)\frac{\chi_1(t)}{\chi_k(t)}.
\eea
\end{subequations}
Integrating (\ref{hannouactive}) with respect to $\br$, 
we also obtain\footnote{Eq. (\ref{hannouactive2}) holds for $\kp \ge1$,
while (\ref{reakinet2}) holds for $k\ge 1$.
}
\bea
w^{(h)}_{k,\kp}(t)=
& &-\!\int \!d\br ~\beta_k(r)F_{k,\kp}(\br,t)+p_{k-1}(t)\nu_{1,\kp}(t) \non\\
& &-B_k(t)\nu_{k,\kp}(t)+B_{k+1}(t)\nu_{k+1,\kp}(t) \non\\
& &-P_k(t)\nu_{k,\kp}(t)+P_{k-1}(t)\nu_{k-1,\kp}(t)
~~~~(\mbox{for}~k\ge 2).
\label{hannouactive2}
\eea
According to the similar procedure,
reaction terms regarding the tail of the ($k,\kp$)-chain
are derived as follows:
\bea
W^{(t)}_{k,\kp}(\br,t)=
& &-\beta_{\kp}(r)F_{k,\kp}(\br,t)+p_{\kp-1}(t)F_{k,1}(\br,t) \non\\
& &-B_{\kp}(t)F_{k,\kp}(\br,t)+B_{\kp+1}(t)F_{k,\kp+1}(\br,t) \non\\
& &-P_{\kp}(t)F_{k,\kp}(\br,t)+P_{\kp-1}(t)F_{k,\kp-1}(\br,t)
~~~~(\mbox{for}~k,\kp\ge 2) 
\label{hannouactive22}
\eea
and
\bea
w^{(t)}_{k,\kp}(t)=& &
-\!\int \!d\br~\beta_{\kp}(r)F_{k,\kp}(\br,t)+p_{\kp-1}(t)\nu_{k,1}(t) \non\\
& &-B_{\kp}(t)\nu_{k,\kp}(t)+B_{\kp+1}(t)\nu_{k,\kp+1}(t) \non\\
& &-P_{\kp}(t)\nu_{k,\kp}(t)+P_{\kp-1}(t)\nu_{k,\kp-1}(t)
~~~~(\mbox{for}~\kp\ge 2).
\label{reakinet2}
\eea


Next, we derive the kinetic equation for dangling 
chains, i.e.,
eq. (\ref{kineticeq}) with $k=1,\kp\ge 2$ 
or $k\ge 2,\kp=1$.
The reaction term $w^{(h)}_{1,\kp}(t)$ 
can be derived as follows.
The ($1,\kp$)-chain is generated by detaching a head of
the ($l,\kp$)-chain ($l\ge 3$) from the $l$-junction.
The increment of $\nu_{1,\kp}$ due to this process per unit time is
\bea
\sum_{l\ge 3}\int d\br~\beta_l(r)F_{l,\kp}(\br,t).
\label{dan1no1}
\eea
The number of the $(1,\kp)$-chain 
also increases when the 2-junction to which
a head of the $(2,\kp)$-chain is belonging divides into two unreacted groups.
The increment according to this procedure is given by (see Appendix \ref{appen1})
\bea
\int d\br~\beta_2(r)F_{2,\kp}(\br,t)
+\left( \int d\br~\beta_2(r)f_2(\br,t) \right)\nu_{2,\kp}(t).
\label{dan2no2}
\eea
On the contrary, $(1,\kp)$-chains are annihilated if the head of
the $(1,\kp)$-chain is connected with $l$-junction ($l\ge 2$).
The decrement of $\nu_{1,\kp}(t)$ caused by this process is given by
$\Bigl( \sum_{l\ge 2} p_l(t) \Bigr)\nu_{1,\kp}(t)$.
Besides, $(1,\kp)$-chains are annihilated when its head captures the unreacted
group.
The decrement due to this reaction is estimated to be $2p_1(t)\nu_{1,\kp}(t)$ 
(see Fig.\ref{hitots5} in Appendix \ref{appen1} as a reference)
As a result, we obtain the reaction term with respect to the head of 
($1,\kp$)-chain as
\footnote{Eq. (\ref{dankore45423}) holds for any $\kp(\ge1)$
and (\ref{dankore4542}) is satisfied for any $k(\ge1)$.}
\bea
w_{1,\kp}^{(h)}(t)=
& &\sum_{l\ge 2}\int d\br~\beta_l(r)F_{l,\kp}(\br,t)
-\Bigl( \sum_{l\ge 1} p_l(t) \Bigr)\nu_{1,\kp}(t) \non\\
& &+B_2(t)\nu_{2,\kp}(t)-p_1(t)\nu_{1,\kp}(t).
\label{dankore45423}
\eea
The reaction term $w_{1,\kp}^{(t)}$ regarding the tail of ($1,\kp$)-chain 
($\kp\ge 2$) is given by (\ref{reakinet2}) with $k=1$.
According to the similar way, the reaction term for the tail of $(k,1)$-chains 
is found to be
\bea
w_{k,1}^{(t)}(t)=
& &\sum_{l\ge 2}\int d\br~\beta_l(r)F_{k,l}(\br,t)
-\Bigl( \sum_{l\ge 1} p_l(t) \Bigr)\nu_{k,1}(t) \non\\
& &+B_2(t)\nu_{k,2}(t)-p_1(t)\nu_{k,1}(t)
\label{dankore4542}
\eea
and 
$w_{k,1}^{(h)}$ ($k\ge 2$) is given by (\ref{reakinet2}) with $\kp=1$.
Finally, the kinetic equation for isolated chain can be obtained 
by setting $k=\kp=1$ in (\ref{kineticeq}), where
$w_{1,1}^{(h)}(t)$ (or $w_{1,1}^{(t)}(t)$) 
is given by (\ref{dankore45423}) with $\kp=1$ (or $k=1$).

By making use of (\ref{hannouactive2}), (\ref{reakinet2}), (\ref{dankore45423}) 
and (\ref{dankore4542}),
we can also obtain the equation for $\chi_k^{(\cdot)}(t)$ 
as follows:
\begin{subequations}\label{kinechi2}
\bea
\frac{d\chi_k^{(\cdot)}(t)}{dt}=
& &-\!\left(\int \!d\br ~\beta_k(r)f_{k}^{(\cdot)}(\br,t)\right)
\chi_k^{(\cdot)}(t)
+p_{k-1}(t)\chi_{1}^{(\cdot)}(t) \non\\
& &-B_k(t)\chi_{k}^{(\cdot)}(t)+B_{k+1}(t)\chi^{(\cdot)}_{k+1}(t) \non\\
& &-P_k(t)\chi_{k}^{(\cdot)}(t)+P_{k-1}(t)\chi_{k-1}^{(\cdot)}(t)
~~~~(\mbox{for}~k\ge 2), 
\label{kinechi2222}
\\
\frac{d\chi_1^{(\cdot)}(t)}{dt}=
& &\sum_{l\ge 2}\left(\int d\br~\beta_l(r)f^{(\cdot)}_{l}(\br,t)\right)
\chi_{l}^{(\cdot)}(t)
-\Bigl( \sum_{l\ge 1} p_l(t) \Bigr)\chi^{(\cdot)}_{1} \non\\
& &+B_2(t)\chi_{2}^{(\cdot)}(t)
-p_1(t)\chi^{(\cdot)}_{1}(t).
\label{kinechi1}
\eea
\end{subequations}
One can confirm 
from
(\ref{kinechi2})
that the total number of chains conserves, i.e.,
\bea
\frac{d}{dt}\sum_{k\ge 1}\sum_{\kp\ge 1}\nu_{k,\kp}(t)
=\sum_{k\ge 1}\frac{d\chi_{k}^{(h)}(t)}{dt}
=\sum_{\kp\ge 1}\frac{d\chi_{\kp}^{(t)}(t)}{dt}
=0. \label{hozonn}
\eea
Note that (\ref{hozonn}) holds for arbitrary 
$\beta_k(r)$ and $p_k$.
Hereafter, we denote the number of total chains (per unit volume) as $n$, i.e.,
\bea
n\equiv\sum_{k\ge 1}\sum_{\kp\ge 1}\nu_{k,\kp}(t). \label{conmas}
\eea


Up to now, the head and the tail of
the chain have been distinguished
for convenience.
Since we are treating 
symmetric chains 
actually, however, 
subscript of $\nu_{k,\kp}$ 
is exchangeable:
$\nu_{k,\kp}(t)=\nu_{k,\kp}(t)$,
and 
hence kinetic equations for $(k,\kp)$-chain (\ref{kineticeq})
with the reaction term
(\ref{hannouactive2}), (\ref{reakinet2}), (\ref{dankore45423}) and (\ref{dankore4542})
can be summarized into
\bea
\frac{d\nu_{k,\kp}(t)}{dt}=w_{k,\kp}(t)+w_{\kp,k}(t)
\label{kineeqsaigo}
\eea
where, for $\kp\ge 1$,
\begin{subequations}\label{kinekazu2}
\bea
& &w_{k,\kp}(t)= 
-\int d\br \beta_k(r)F_{k,\kp}(\br,t)+p_{k-1}(t)\nu_{1,\kp}(t) 
-(B_k(t)+P_k(t))\nu_{k,\kp}(t) \non\\
& &~~~~~~~~~~~~~+B_{k+1}(t)\nu_{k+1,\kp}(t) +P_{k-1}(t)\nu_{k-1,\kp}(t)
~~~~(\mbox{for}~~k\ge 2), \\
& &w_{1,\kp}(t)=
\sum_{l\ge 2}\int d\br~\beta_l(r)F_{l,\kp}(\br,t)
+B_2(t)\nu_{2,\kp}(t)
-\left(p_1(t)+\sum_{l\ge 1} p_l(t)\right)\nu_{1,\kp}(t). 
\hspace*{1cm}
\eea
\end{subequations}
Further, we find that
$\chi_k(t)=\chi_k^{(h)}(t)=\chi_k^{(t)}(t)$
satisfies the following equation:
\bea
\frac{d\chi_k(t)}{dt}=u_k(t),
\label{chiequ2}
\eea
where
\begin{subequations}\label{kinechi200mato}
\bea
u_k(t)
=& &-k\bra \beta_k(r)\ket(t)\chi_k(t)
+k\bra \beta_{k+1}(r)\ket(t)\chi_{k+1}(t) 
\non\\
& &
+kp_{k-1}(t)\chi_1(t)-kp_k(t)\chi_1(t) ~~~(\mbox{for}~k\ge 2), 
\label{kinechi200}\\
u_1(t)
=& &\sum_{l\ge 2}\bra \beta_{l}(r)\ket(t)\chi_{l}(t)
-\Bigl(\sum_{l\ge 1}p_l(t)\Bigr)\chi_1(t)
 \non\\
& &+\bra \beta_2(r)\ket(t)\chi_2(t)-p_1(t)\chi_1(t).
\label{kinechi100}
\eea
\end{subequations}
Evolution equation (\ref{activeeq}) with the reaction terms
given by (\ref{hannouactive}) and (\ref{hannouactive22})
as well as the kinetic equations 
for ($k,\kp$)-chain (\ref{kineeqsaigo}) together with (\ref{kinekazu2})
and for $k$-chain (\ref{chiequ2}) together with (\ref{kinechi200mato})
are fundamental equations 
of our transient networks
and will be solved in the following sections.

\section{Reaction Rates of Functional Groups}
\label{secrates}

In the rest of this article,  
we treat the dissociation rate of the functional group
as constant independent of the end-to-end length
of a chain. 
This treatment is valid since we are going to 
apply a small oscillatory deformation
to the system
and the change of the dissociation rate through $r$ is quite small
in this case.
Then, according to TE,
the dissociation rate 
from the $k$-junction
takes a form:
$\beta_k=\omega_0\exp\left(-W_k/k_BT\right)$
\cite{tanaed2}
where $\omega_0$ is a reciprocal of a microscopic time
and $W_k$ is a potential barrier for 
the dissociation.
We further assume that the potential barrier 
does not depend on the multiplicity of the junction and set $W_k=W$ for all $k$.
Then the dissociation rate also does not depend on the junction multiplicity
and is written as
\bea
\beta_k=\omega_0\exp\left(-W/k_BT\right)~(\equiv\beta).
\label{disodef}
\eea

\begin{figure}[t]
\begin{center}
\includegraphics*[scale=0.6]{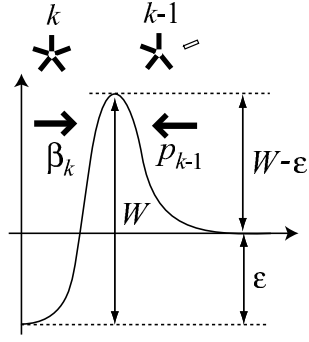}
\end{center}
\vspace*{-0.5cm}
\caption{The potential barrier around a $k$-junction
for the dissociation-association reaction
of the functional group. 
}
\label{poten}
\end{figure}

The connection rate of an unreacted group to a $k$-junction 
should increase with increasing the number of functional groups 
forming the $k$-junction in the close vicinity of the unreacted group.
We assume that the connection rate takes the form
\bea
p_k(t)=\omega_0\exp\left(-(W-\epsilon)/k_BT\right)k\mu_k(t)v_0h_k, \label{psaiso0}
\eea
where $\epsilon$ is a binding energy between the functional group
and the junction\footnote{It is also assumed 
that the binding energy
does not depend on the junction multiplicity.} (see Fig.\ref{poten}),
$k\mu_k(t)v_0$ 
is the number of functional groups 
forming $k$-junctions 
in the effective volume $v_0$ of 
the (unreacted) functional group,
and $h_k$ is a proportional factor given in the next section.
Note that the connection rate depends on time through the number of $k$-junction
in general.
Eq. (\ref{psaiso0}) can be rewritten as
\bea
p_k(t)&=&\beta\lambda(T)\psi q_k(t)h_k
\label{qdefkore}
\eea
where
\bea
q_k(t)\equiv\frac{k\mu_k(t)}{2n}=\frac{\chi_k(t)}{n}
\label{coneodef}
\eea
is the probability that an arbitrary chosen functional group belongs to a $k$-junction,
$\psi\equiv 2nv_0$ is the volume fraction of functional groups, and
\bea
\lambda(T)\equiv\exp\left(\frac{\epsilon}{k_BT}\right)
\eea
is the association constant of the reactions
introduced in ref. \cite{tanastock}.
Thus the connection rate to 
the $k$-junction
is proportional to the volume fraction $\psi q_k$
of $k$-junctions.

\section{Equilibrium Properties}
\label{eqiv}

In equilibrium state, the number $\chi_k$ of $k$-chain, or equivalently $q_k$, 
is formally
obtained 
by setting the left-hand side of (\ref{kinechi2}) to 0.
(Here and hereafter, all quantities in equilibrium
are denoted 
without the argument $t$. 
For example, $\nu_{k,\kp}$ is the number of $(k,\kp)$-chains in equilibrium.)
Thus we find
\bea
~~~~~~~~~q_k=\frac{p_{k-1}}{\bra \beta_k(r)\ket}q_1 ~~~
(\mbox{for}~k\ge 2)
\label{motopha}
\eea
where $q_1$ is obtained from the normalization condition 
$\sum_{k\ge 1}q_k=1$:\footnote{This equation is equivalent to the
number conservation equation $\sum_{k\ge 1}\chi_k=n$ for chains.}
\bea
q_1=\frac{1}{1+\sum_{k\ge 2}p_{k-1}/\bra\beta_k(r)\ket}.
\eea
On substitution of 
(\ref{disodef}) and (\ref{qdefkore}),
eq. (\ref{motopha}) becomes
\bea
~~~~~~~~~~~
q_k=\lambda\psi h_{k-1} q_{k-1}q_1
~~~~(\mbox{for}~k\ge 2). \label{seconreac}
\eea
By an iterating procedure, we find the following mass action law:
\bea
& &q_k=\gamma_k(\lambda\psi)^{k-1}q_1^k
~~~(\mbox{for}~k\ge 2), \label{equivk}\\
& &q_1=1/\gamma(z), \label{equiv1}
\eea
where 
\bea
& &\gamma_k\equiv \left\{
\begin{array}{@{\,}ll}
1                    &  (k=1) \\
\prod_{l=1}^{k-1}h_l & (k\ge2) 
\end{array}
\right., \label{gammateigisd}\\
& &\gamma(z)\equiv\sum_{k\ge 1}\gamma_kz^{k-1}
\eea
and $z\equiv\lambda\psi q_1$.
If $\lambda$ and $\psi$ are given,
$q_1$ can be derived by solving (\ref{equiv1}).
Then we can obtain $q_k~(k\ge 2)$ from (\ref{equivk}).
The association condition (\ref{equivk}) 
together with (\ref{equiv1}) 
have been derived 
in the theory of 
thermoreversible gelation with junctions of variable multiplicity
developed 
by Tanaka and Stockmayer (referred to as TS)
\cite{tanastock}.\footnote{In 
this series of articles, we denote
the probability that an arbitrary chosen functional group belongs to 
a $k$-junction as $q_k$,
although the same quantity is represented as $p_k$ in ref.\cite{tanastock},
because the symbol $p_k$ is used as the connection rate of functional groups.
}
In the TS theory, $\gamma_k$ is interpreted as a factor giving the surface correction 
for the binding energy,
although it is set to unity for all $k$ 
for simplicity.
We will adjust $h_k$ (and hence $\gamma_k$) 
to derive
specific models for junctions (see below).
TS has shown that
most quantities describing transient gels in equilibrium
depend on the polymer 
volume fraction $\psi$
through
the combination of $\lambda(T)$ and $\psi(=2\phi/N)$.
This holds not only in equilibrium state but also
under small deformation as shown in the next section.
Therefore, we use 
$c\equiv \lambda(T)\psi$ 
as the reduced polymer concentration 
in the following.
In equilibrium, (\ref{kinekazu2}) reduces to a simpler form:
\begin{subequations}\label{kinekazu211}
\bea
& &w_{k,\kp}=	
-\beta k(1+zh_k)\nu_{k,\kp}
+\beta k\nu_{k+1,\kp}
+\beta (k-1)zh_{k-1}\nu_{k-1,\kp}
+\beta c h_{k-1}q_{k-1}\nu_{1,\kp} \non\\
& &\hspace*{11.5cm}(\mbox{for}~k\ge 2), \\
& &w_{1,\kp}=
\beta\left( \nu_{2,\kp}+\sum_{l\ge 2}\nu_{l,\kp}\right)
-\beta c\biggl(h_1q_1+\sum_{l\ge 1}h_lq_l\biggr)\nu_{1,\kp}.
\eea
\end{subequations}
By solving a equation $(d\nu_{k,\kp}/dt\!=)~w_{k,\kp}+w_{\kp,k}=0$, 
we can obtain the number of $(k,\kp)$-chains 
in equilibrium as
\bea
\nu_{k,\kp}=nq_kq_{\kp}.
\label{equnukk}
\eea
Note that $\nu_{k,\kp}/n=q_kq_{\kp}$ is the probability 
that 
one end of an arbitrary chosen chain is incorporated into a $k$-junction
while its other end is belonging to a $\kp$-junction.
Thus (\ref{equnukk}) is compatible with the definition of the ($k,\kp$)-chain.


Now we consider two special cases as for the multiplicity that the junction can take; 
1) a saturating junction model and 2) a fixed multiplicity model.
These two models have been considered by Tanaka and Stockmayer \cite{tanastock}
in studies of the phase behavior of associating polymer solutions 
in equilibrium.\footnote{They have also considered 
a minimum junction model \cite{tanastock}, 
where junctions are allowed to take the multiplicity 
$k=1$ (unreacted) and $k=s_0,s_0+1,\cdots$.}
In the saturating junction model, 
junction multiplicity has an upper limit $s_m$, that is,
the multiplicity is allowed to take a limited range $k=1,2,...s_m$.
This model might be applied to the junction
formed by ionic dipolar interaction or hydrophobic aggregation
because the space around such a junction is packed with polymer chains
and steric hindrance among 
them restricts the number of chains connected with the junction.
\cite{tanastock}.
The mean multiplicity generally depends on the reduced polymer concentration 
in this model.
On the other hand, in the fixed multiplicity model,
the multiplicity can take only one fixed number $s$, i.e.,
we have only $k=1$ (unreacted) and $k=s$ (reacted) irrespective of the 
reduced polymer concentration.

\subsection{Saturating Junction Model}
\label{satuequ}

\begin{figure}[t]
\begin{center}
\includegraphics*[scale=0.4]{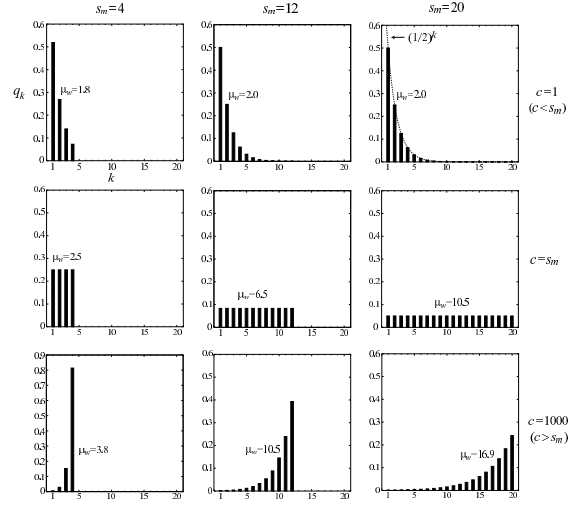}
\end{center}
\vspace*{-0.5cm}
\caption{
The probability distribution $q_k$ 
that a randomly selected functional group to be in a $k$-junction 
for the maximum multiplicity 
$s_m=4$ (left column), $12$ (middle column), $20$ (right column),
and for the reduced polymer concentration 
$c=1$ (top raw),
$c=s_m$ (middle row),
$c=1000$ (bottom row).
The weight-average multiplicity $\mu_w$ of the junction is shown in each figure.
}
\label{chi2}
\end{figure}

We can impose the upper limit on the junction multiplicity
by assuming that $h_k$ is given as a stepwise function:
\bea
& &h_k= \left\{
\begin{array}{@{\,}ll}
1 & (1\le k \le s_m-1) \\
0 & (k\ge s_m)
\end{array}
\right.. \label{satuhk}
\eea
For such $h_k$, (\ref{gammateigisd}) reduces to
\bea
& &\gamma_k=
\left\{
\begin{array}{@{\,}ll}
1 & (1\le k\le s_m) \\
0 & (k\!\ge\! s_m+1)
\end{array}
\right.,
\eea
and hence $q_k$ takes the form
\bea
& &q_k= \left\{
\begin{array}{@{\,}ll}
(cq_1)^{k-1}q_1 & (2\le k\le s_m) \\
0 & (k \ge s_m+1)
\end{array}
\right..
\label{qkteiji}
\eea
Thus, junctions with the multiplicity larger than $s_m$ do not exist anymore.
We can obtain $q_1$ by solving (\ref{equiv1}):
\bea
\frac{1}{q_1}
=\frac{1-(cq_1)^{s_m}}{1-cq_1}.
\label{deriveq1e}
\eea
Note that the right-hand side of (\ref{deriveq1e})
(denoted as $g(q_1)$ for simplicity)
is $s_m$ when $q_1=1/c$.
Therefore, when the condition $s_m=c$ is satisfied,
a solution of (\ref{deriveq1e}) is $q_1=1/c$.
As a result, $q_k$ does not depend on $k$, i.e., $q_k=1/c$ for all $k(\le s_m)$
(see middle row figures of Fig.\ref{chi2}).
When $s_m$ is larger than $c$, $g(q_1=1/c)=s_m$ is also larger than $c$.
Therefore, 
a solution of (\ref{deriveq1e}) satisfies a condition $q_1<1/c$
because $g(q_1)$ is an increasing function with respect to $q_1$.
Consequently, $q_k$ is a decreasing function with respect to $k$
(see top raw figures of Fig.\ref{chi2}).
On the contrary, 
when $s_m$ is smaller than $c$, 
a solution of (\ref{deriveq1e}) fulfills a condition $q_1>1/c$, 
and hence $q_k$ is an increasing function 
of $k$ (see bottom row figures of Fig.\ref{chi2}).

\begin{figure}[t]
\begin{center}
\includegraphics*[scale=0.5]{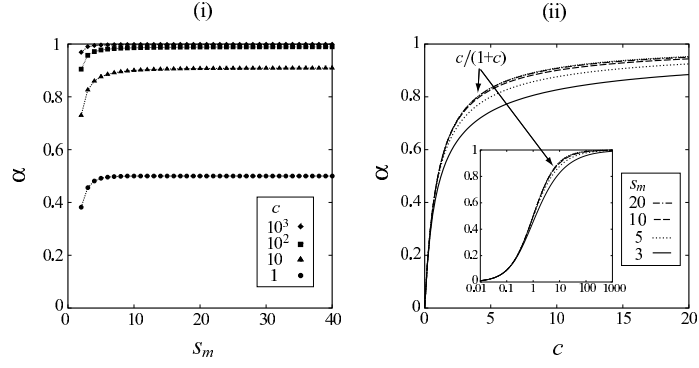}
\end{center}
\vspace*{-0.5cm}
\caption{
The extent of reaction $\alpha$ of the saturating junction model
as a function of the maximum multiplicity $s_m$ (i),
and of the reduced polymer concentration $c$ (ii).
The reduced concentration is varying from curve to curve in (i), 
while the maximum multiplicity is changing 
in (ii).
The inset of (ii) shows the linear-log plot of $\alpha$ as a function of $c$.
Dotted curves (behind the curves for $s_m=20$) 
in (ii) represent $\alpha=c/(1+c)$.
}
\label{alpha2}
\end{figure}

Fig.\ref{alpha2} (i) shows the extent of reaction $\alpha=1-q_1$ 
as a function of $s_m$ for several reduced concentration $c$.
We see that $\alpha$ approaches to a fixed value for each $c$
as $s_m$ increases.
This value can be estimated as follows.
In an extreme case 
where $s_m$ is much larger than $c$,
the right-hand side of (\ref{deriveq1e}) is close to $1/(1-cq_1)$ 
thanks to the condition $cq_1<1$.
Thus we find
\bea
q_1= \frac{1}{1+c}~~~(\mbox{for}~s_m\gg c),
\label{11111111}
\eea
and therefore
\bea
\alpha= \frac{c}{1+c}~~~(\mbox{for}~s_m\gg c).
\label{11111112}
\eea
In this extreme, $q_k$ 
is expressed as
\bea
q_k=\frac{1}{c}\left(\frac{c}{1+c}\right)^k=\frac{1}{c}\exp[-k/\kappa]
~~~(\mbox{for}~s_m\gg c),
\label{11114324}
\eea
where
\bea
\kappa\equiv 1/\log[(1+c)/c]
\eea
indicates the width of the distribution.
As an example,
(\ref{11114324}) is drawn in Fig.\ref{chi2}
for $c=1$ and $s_m=20$.
As seen from Fig.\ref{alpha2} (ii),
the extent of reaction behaves as $c/(1+c)$ for $c$ much smaller than $s_m$,
while it approaches to 1 if $c$ exceeds $s_m$.

\begin{figure}[t]
\begin{center}
\includegraphics*[scale=0.5]{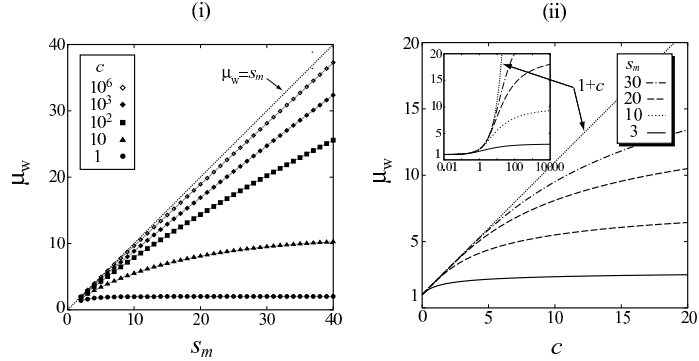}
\end{center}
\vspace*{-0.5cm}
\caption{
The weight-average multiplicity $\mu_w$ of the junction
for the saturating junction model
as a function of the maximum multiplicity $s_m$ (i),
and of the reduced polymer concentration $c$ (ii).
The reduced concentration is varying from curve to curve in (i), 
while the maximum multiplicity is changing 
in (ii).
The inset of (ii) shows the linear-log plot of $\mu_w$
as a function of $c$.
}
\label{alpha1}
\end{figure}

The weight-average multiplicity defined by
\bea
\mu_w=\sum_{k= 1}^{s_m}k q_k
\eea
is shown in Fig.\ref{chi2}
for each $s_m$ and $c$ and is depicted in Fig.\ref{alpha1}
as a function of $s_m$ (i) and of $c$ (ii).
Note that $\mu_w$ includes unreacted groups as $k=1$ junctions.
When $s_m$ is much larger than $c$,
it is close to (see Fig.\ref{alpha1} (i) and (ii))
\bea
\mu_w=1+c~~(\mbox{for}~s_m\gg c).
\eea
When the reduced concentration is low and
satisfies the condition $c\ll s_m$, 
we see from top three figures of Fig.\ref{chi2} that
$\mu_w$ is much smaller than $s_m$ and close to 1 irrespective the value of $s_m$.
This is because there are many unreacted groups (represented by $q_1$) 
in the system.
On the other hand, when the reduced concentration is high and $c\gg s_m$, 
$\mu_w$ is close to $s_m$ (see bottom three figures of Fig.\ref{chi2})
since many functional groups are incorporated into
junctions with $k=s_m$.
These tendencies can also be confirmed from Fig.\ref{alpha1} (i).
When $s_m$ is much smaller than $c$, 
then $\mu_w$ is approximately $s_m$,
while $\mu_w$ approaches to $1+c$ when $s_m$ exceeds $c$ for each $c$.
As seen from Fig.\ref{alpha1} (ii), 
$\mu_w$ behaves as $1+c$ 
irrespective of the value of $s_m$ when $c\ll s_m$,
and it approaches to $s_m$ as $c$ increases beyond $s_m$.

\subsection{Fixed Multiplicity Model}

Let $h_k$ be 
\bea
h_k= \left\{
\begin{array}{@{\,}ll}
\delta          & (1\le k< s-1) \\
\delta^{-(s-2)} & (k=s-1) \\
0 & (k> s-1)
\end{array}
\right.,\label{deltasyou}
\eea
where $\delta$ is a positive value.
Then
$q_k$ becomes
\bea
q_k&=& \left\{
\begin{array}{@{\,}ll}
(\delta c)^{k-1}q_1^k & (1\le k < s) \\
c^{s-1}q_1^s & (k=s) \\
0 & (k>s)
\end{array}
\right..
\eea
All junctions take approximately the same multiplicity $s$
if we employ $\delta$ much smaller than 1\footnote{On the other hand,
the saturating junction model 
can be realized by setting $\delta=1$ in (\ref{deltasyou}).}
because $q_k$ is nearly equal to 0 except for the case of $k=s$ (and $k=1$), i.e.,
\bea
q_k&\simeq&\left\{
\begin{array}{@{\,}ll}
c^{s-1}q_1^s & (k=s) \\
0 & (k\neq s)
\end{array}
\right. ~~(\mbox{for}~\delta\ll 1). 
\label{d0.012eq}
\eea
It should be emphasized here that we cannot 
fix the junction multiplicity rigorously at $s$ by putting $\delta=0$
because 
junctions whose multiplicity is less than $s$ must exist
for the creation of $s$-junctions 
under the assumption of stepwise reactions.
We set $\delta=0.01$ in this series of papers.
This $\delta$ value
is enough to describe characteristic properties of the system 
where junctions can take only one
multiplicity.
In the following, 
we often use the equal sign
instead of the nearly equal sign ($\simeq$)
for equations  (such as (\ref{d0.012eq}))
that approximately hold for small $\delta$.
In the fixed multiplicity model, 
the extent of reaction is given by $\alpha=q_s$
because of the normalization condition $q_1+q_s=1$.


\begin{figure}[t]
\begin{center}
\includegraphics*[scale=0.45]{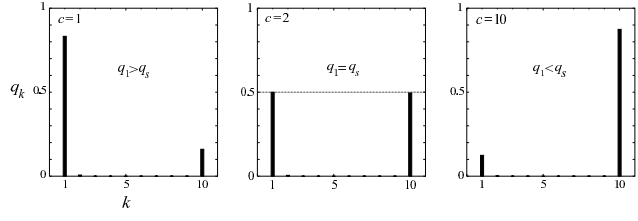}
\end{center}
\vspace*{-0.5cm}
\caption{
The probability distribution $q_k$ 
that a randomly selected functional group to be in a $k$-junction 
for the reduced concentration $c=1$ (left), $2$ (middle), and $10$ (right).
The multiplicity of the junction is fixed at $s=10$.
}
\label{chi10}
\end{figure}

The probability $q_1$ to find an unreacted group can be obtained
by solving (\ref{equiv1}):
\bea
\frac{1}{q_1}=1+(cq_1)^{s-1}.
\label{fixq1der}
\eea
The right-hand side of (\ref{fixq1der}) 
(denoted as $g(q_1)$) is equal to 2 for $q_1=1/c$.
Therefore, if $c=2$, 
a solution of (\ref{fixq1der}) is given by $q_1=1/2(=q_s)$.
If $c>2$,
a solution of (\ref{fixq1der}) satisfies a condition
$1/c<q_1<1/2$ because $g(q_1=1/2)=1+(c/2)^{(s-1)}$ is greater than $2$.
Then $q_s(>1/2)$ is larger than $q_1$ (see Fig.\ref{chi10}) indicating that
there are more reacted groups than unreacted ones.
Note that when $s$ is much larger than 1, 
$g(q_1)$  is equal to 1 for $q_1\le 1/c$ but it almost diverges at $q_1>1/c$,
and therefore, a solution of (\ref{fixq1der}) approaches to $1/c$.
On the other hand, if $c<2$, the solution of (\ref{fixq1der}) satisfies 
a condition $1/2<q_1<\min(1,1/c)$, 
and it approaches to $1/c$ for $c$ satisfying $1<c<2$, or to $1$ for $c\le1$ 
with increasing $s$.
Summarizing,
\bea
q_1=\left\{
\begin{array}{@{\,}ll}
1/c & (c>1)  \\
1 & (c\le1)  
\end{array}
\right. ~~(\mbox{for}~s\gg1)
\eea
or, equivalently,
\bea
\alpha=\left\{
\begin{array}{@{\,}ll}
1-1/c & (c>1)  \\
0 & (c\le1)
\end{array}
\right. ~~(\mbox{for}~s\gg1).
\eea
Thus junctions suddenly appear at $c=1$ in this extreme.

\begin{figure}[t]
\begin{center}
\includegraphics*[scale=0.5]{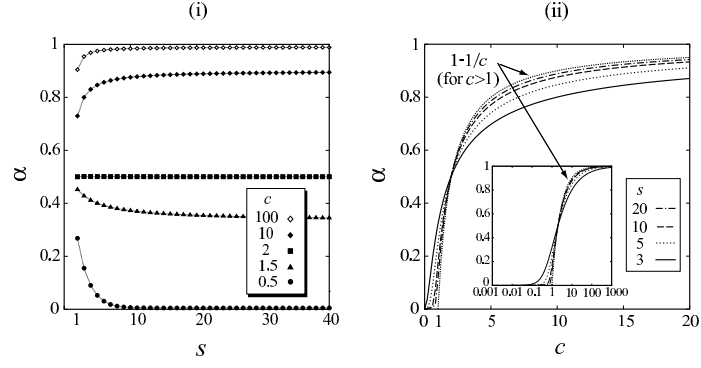}
\end{center}
\vspace*{-0.5cm}
\caption{
The extent of reaction 
of the fixed multiplicity model
as a function of the reduced polymer concentration $c$ (i),
and of the junction multiplicity $s$ (ii).
The inset of (ii) shows the linear-log plot of $\alpha$
as a function of $c$.
%
}
\label{alpha4}
\end{figure}

Fig.\ref{alpha4} (i) shows the extent of reaction as a function of $s$.
We can confirm the tendency stated above.
If $c\le1$, the extent of reaction approaches to 0 with increasing $s$.
If $c>1$, on the other hand, it approaches to $1-1/c$ for a given $c$.
But when $c=2$, it does not depend on $s$.
Fig.\ref{alpha4} (ii) shows the extent of reaction as a function of $c$.
It increases abruptly around $c=1$ when $s$ is large.
Such a sharp increase in $\alpha$ 
stems from the fact that the junctions can take (approximately) 
only one multiplicity;
even if several functional groups 
spend a certain duration of time in the close vicinity 
of each other,
they cannot aggregate unless $s$ groups 
participate in this event.
Actually, this tendency cannot be observed
in the saturating junction model (see Fig.\ref{alpha2})
since the junction can be formed by any number of functional groups
less than a certain value.

\section{Dynamic-Mechanical and Viscoelastic Properties}
\label{secv}

Now, we apply a small oscillatory shear deformation to the 
present system whose rate of deformation tensor is represented by
\bea
\hat{\kappa}(t)=\left(
\begin{array}{ccc}
0 & \epsilon\omega\cos \omega t & 0 \\
0 & 0 & 0 \\
0 & 0 & 0 
\end{array}
\right), \label{zurihenkeika}
\eea
where $\epsilon$ is a dimensionless infinitesimal amplitude
and $\omega$ is the frequency of the oscillation.
By substituting (\ref{zurihenkeika}) 
into the left-hand side of (\ref{activeeq}),
and by putting (\ref{disodef}) and (\ref{qdefkore}) into the right-hand side
(given by (\ref{hannouactive}) and (\ref{hannouactive22})),
we have the evolution equation for $F_{k,\kp}(\br,t)$ 
given as follows:
\bea
& &\frac{\partial F_{k,\kp}(\br,t)}{\partial t}
+\frac{\partial F_{k,\kp}(\br,t)}{\partial x}\epsilon  y\omega\cos\omega t \non\\
& &\hspace{1.5cm}=-Q_{k,\kp}(t)F_{k,\kp}(\br,t) 
+B_{k+1}F_{k+1,\kp}(\br,t)
+B_{\kp+1}F_{k,\kp+1}(\br,t) 
\non\\
& &\hspace{2cm}+P_{k-1}(t)F_{k-1,\kp}(\br,t)
+P_{\kp-1}(t)F_{k,\kp-1}(\br,t) 
\non\\
& &\hspace{2cm}+\beta c\bigl(h_{k-1}q_{k-1}(t)\nu_{\kp,1}(t)
+h_{\kp-1}p_{\kp-1}(t)\nu_{k,1}(t)\bigr)f_0(\br)
~~(\mbox{for}~k,\kp\ge 2),
\eea
where\footnote{$z(t)\equiv\lambda\psi q_1(t)$.}
\bea
& &B_k=\beta(k-1), \label{bknew} \\
& &P_k(t)=\beta z(t)kh_k \label{pknew}
\eea
as defined in (\ref{BPteigi}),
and
\bea
Q_{k,\kp}(t)\equiv\beta k +P_{k}(t)+\beta\kp +P_{\kp}(t).
\label{teigiq2}
\eea
Here, we expand $F_{k,\kp}(\br,t)$ 
with respect to the powers of $\epsilon$ up to the first order:
\bea
F_{k,\kp}(\br,t)=F^{(0)}_{k,\kp}(\br)+\epsilon F^{(1)}_{k,\kp}(\br,t).
\eea
On the other hand, the first-order terms 
of $\nu_{k,\kp}(t)$ and $q_{k}(t)$
are negligible \cite{tanaed1,intana1},
i.e., they have equilibrium values
and satisfy the association condition (\ref{equivk})
even under the small shear deformation.
The zeroth-order term of $F_{k,\kp}(\br)$ represents its
equilibrium value in the absence of the oscillation, 
and hence it is written as 
\bea
F^{(0)}_{k,\kp}(\br)=\nu_{k,\kp}f_0(\br).
\eea
By comparing the order, we see that the first-order term
$F^{(1)}_{k,\kp}(\br,t)$ satisfies the following equation:
\bea
& &\frac{\partial F_{k,\kp}^{(1)}(\br,t)}{\partial t}
-\nu_{k,\kp}\frac{3xy}{Na^2}f_0(\br)\omega\cos\omega t \non\\
& &\hspace{1.5cm}=-Q_{k,\kp}F_{k,\kp}^{(1)}(\br,t) 
+B_{k+1}F^{(1)}_{k+1,\kp}(\br,t)
+B_{\kp+1}F^{(1)}_{k,\kp+1}(\br,t) 
\non\\
& &\hspace{2cm}+P_{k-1}F^{(1)}_{k-1,\kp}(\br,t)
+P_{\kp-1}F^{(1)}_{k,\kp-1}(\br,t)
~~~~(\mbox{for}~k,\kp\ge 2). 
\label{eqi5i1}
\eea
The solution of (\ref{eqi5i1}) after the long-time limit 
takes the form\footnote{Eq.(\ref{f1teigi3}) assures that
the number of ($k,\kp$)-chains of the first order, say $\nu_{k,\kp}^{(1)}(t)$,
is zero; $\nu_{k,\kp}^{(1)}(t)=\int d\br F^{(1)}_{k,\kp}(\br,t)=0.$}
\bea
F^{(1)}_{k,\kp}(\br,t)=
\Bigl(
g^{\prime}_{k,\kp}(\omega)\sin\omega t+g^{\prime\prime}_{k,\kp}(\omega)\cos\omega t
\Bigl)
\frac{3xy}{Na^2}f_0(\br).
\label{f1teigi3}
\eea
Note that subscripts of 
$g^{\prime}$ (and $g^{\prime\prime}$) are exchangeable, i.e.,
$g_{k,\kp}^{\prime}=g_{\kp,k}^{\prime}$ (or
$g_{k,\kp}^{\prime\prime}=g_{\kp,k}^{\prime\prime}$),
and we can put $g_{k,1}^{\prime}=g_{k,1}^{\prime\prime}\equiv 0$ for $k\ge 1$.
%
Substituting (\ref{f1teigi3}) into (\ref{eqi5i1}),
we obtain the simultaneous equation for 
$g^{\prime}_{k,\kp}$ and 
$g^{\prime\prime}_{k,\kp}$ as follows:
\begin{subequations}\label{kretoek2}
\bea
& &g^{\prime}_{k,\kp}=
\left(-Q_{k,\kp}g^{\prime\prime}_{k,\kp}
+B_{k+1}g^{\prime\prime}_{k+1,\kp}
+B_{\kp+1}g^{\prime\prime}_{k,\kp+1}
+P_{k-1}g^{\prime\prime}_{k-1,\kp}
+P_{\kp-1}g^{\prime\prime}_{k,\kp-1}\right)/\omega+\nu_{k,\kp}, ~~~~~~~~~~~~\\
& &g^{\prime\prime}_{k,\kp}
=\left(Q_{k,\kp}g^{\prime}_{k,\kp}
-B_{k+1}g^{\prime}_{k+1,\kp}
-B_{\kp+1}g^{\prime}_{k,\kp+1}
-P_{k-1}g^{\prime}_{k-1,\kp}
-P_{\kp-1}g^{\prime}_{k,\kp-1}\right)/\omega \\
& &\hspace*{13cm}(\mbox{for~}k,\kp\ge 2) \non
\eea
\end{subequations}
In the high frequency limit, (\ref{kretoek2}) reduces to
\begin{subequations}\label{hjlimits}
\bea
& &g^{\prime}_{k,\kp}(\omega\to\infty)= \nu_{k,\kp}, 
\label{hjlimitsdd} \\
& &g^{\prime\prime}_{k,\kp}(\omega\to\infty)= 0.
\label{kyokugenG}
\eea
\end{subequations}
The shear stress $\sigma_{k,\kp}$ ascribed to the $(k,\kp)$-chain
is derived from the following formula for the Gaussian chain:
\bea
\sigma_{k,\kp}=\frac{3k_BT}{Na^2}\int d\br ~xy F_{k,\kp}(\br,t)
=\epsilon\left[ G_{k,\kp}^{\prime}(\omega)\sin\omega t+
G_{k,\kp}^{\prime\prime}(\omega)\cos\omega t \right]
\eea
where the dynamic shear moduli 
with respect to the $(k,\kp)$-chain are
defined by
\begin{subequations}\label{Gkktotyuu}
\bea
& &G_{k,\kp}^{\prime}(\omega)=k_BT
g^{\prime}_{k,\kp}(\omega),\\
& &G_{k,\kp}^{\prime\prime}(\omega)=k_BT
g^{\prime\prime}_{k,\kp}(\omega).
\eea
\end{subequations}
Since we are assuming that any chains whose both ends are 
associated with
other functional groups
are elastically effective,
the total moduli are given by 
\begin{subequations}\label{koreGpGPP}
\bea
G^{\prime}(\omega)&=&
k_BT\sum_{k\ge 2}\sum_{\kp\ge 2}g^{\prime}_{k,\kp}(\omega), \\
G^{\prime\prime}(\omega)&=&
k_BT\sum_{k\ge 2}\sum_{\kp\ge 2}g^{\prime\prime}_{k,\kp}(\omega).
\eea
\end{subequations}
The high-frequency plateau modulus is then found 
from (\ref{hjlimitsdd}) to be
\bea
G_{\infty}\equiv G^{\prime}(\omega\to\infty)=\nu^{\eff}_0k_BT
\eea
where
\bea
\nu^{eff}_0=\sum_{k\ge 2}\sum_{\kp\ge 2}\nu_{k,\kp}
=n\alpha^2
\label{numTEtyoi}
\eea
is the total number of active chains.

\begin{figure}[t]
\begin{center}
\includegraphics*[scale=0.6]{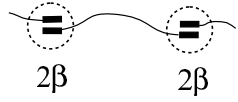}
\end{center}
\vspace*{-0.5cm}
\caption{
The breakage rate of a pairwise junction is $2\beta$ 
because it decays if one of two functional groups in the junction 
is disconnected from the other.
Since active chains are connected with two pairwise junctions
through both ends,
the breakage rate of the active chain is $4\beta$.
}
\label{pair}
\end{figure}

As an example, let us 
study
the simplest case that 
the association is allowed only in pairs.
In other words, only junctions 
whose multiplicity is two are allowed to exist in the system.
There is no difference between two models for junctions 
(i.e., the saturating junction model and fixed multiplicity model)
in this case.
Since we are treating telechelic polymers,
only linearly 
extended
chains can be formed 
rather than a three dimensional network.
Nevertheless, it is worth studying
such simple situation because
this is the only case that
the moduli can be analytically expressed in simple forms,
thereby giving insight into the system with multiple junctions.
Note that $g^{\prime}_{3,2}\!=\!g^{\prime}_{2,3}\!=\!
g^{\prime\prime}_{3,2}\!=\!g^{\prime\prime}_{2,3}\!=\!0$ 
since there are no 3-junctions.
Then (\ref{kretoek2}) reduces to
\begin{subequations}
\bea
& &g^{\prime}_{2,2}
=-Q_{2,2}g^{\prime\prime}_{2,2}/\omega+\nu_{2,2}, \\
& &g^{\prime\prime}_{2,2}=Q_{2,2}g^{\prime}_{2,2}/\omega.
\eea
\end{subequations}
As a result, the moduli becomes the Maxwellian with a single relaxation time 
$1/Q_{2,2}$, i.e.,
\begin{subequations}
\bea
& &G^{\prime}/k_BT=g^{\prime}_{2,2}
=\frac{(Q_{2,2}\omega)^2}{(Q_{2,2}\omega)^2+1}\nu_0^{\eff},\\
& &G^{\prime\prime}/k_BT=g^{\prime\prime}_{2,2}
=\frac{Q_{2,2}\omega}{(Q_{2,2}\omega)^2+1}\nu_0^{\eff},
\eea
\end{subequations}
where the number of active chains is 
$\nu_0^{\eff}=\nu_{2,2}=n(zh_1q_1)^2$,
and the reciprocal of the relaxation time,
or the breakage rate of the active chain,
is given by $Q_{2,2}=4\beta+2P_2=4\beta$.\footnote{The probability $p_2$ 
that an unreacted group is connected with a 2-junction is zero
since 3-junction is not allowed to exist by definition. 
Therefore, $P_2\propto p_2=0$.}
The coefficient 4 of the chain breakage rate
stems from the annihilation rate
$2\beta$ of the junction to which one end of the chain is belonging
and $2\beta$ for the other end (see Fig.\ref{pair}).

\subsection{Saturating Junction Model}

\begin{figure}[t]
\begin{center}
\includegraphics*[scale=0.5]{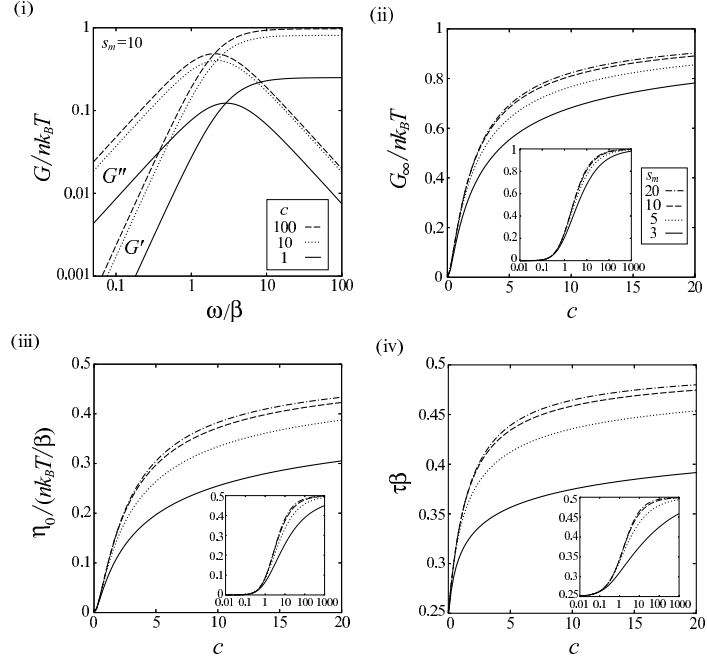}
\end{center}
\vspace*{-0.5cm}
\caption{
(i) The dynamic shear moduli (reduced by $nk_BT$) 
for the saturating junction model
as a function of the frequency.
The reduced polymer concentration $c$ is varying from curve to curve,
while the maximum multiplicity is fixed at $s_m=10$.
(ii) The reduced plateau modulus, 
(iii) reduced zero-shear viscosity, 
and (iv) relaxation time
plotted against the reduced concentration.
The maximum multiplicity is varying from curve to curve.
The insets of (ii) $\sim$ (iv) show the linear-log plot of each quantity
as a function of $c$.
}
\label{getatau1}
\end{figure}

Now we study dynamic-mechanical and viscoelastic properties of 
the system
where multiple junctions (i.e., $k>2$) are allowed to exist.
We first consider the saturating junction model, i.e., 
the junction takes a limited range $k=1,2,...s_m$ of the multiplicity.
This condition can be attained by employing (\ref{satuhk}) for $P_k$, i.e.,
\bea
& &P_k
=\left\{
\begin{array}{@{\,}ll}
\beta zk & (1\le k\le s_m-1) \\
0 & (k \ge s_m)
\end{array}
\right.. \label{newPk2}
\eea
By putting (\ref{newPk2}) 
into (\ref{kretoek2}) and by solving a set of linear algebraic equations for 
$g^{\prime}_{k,\kp}$ and $g^{\prime\prime}_{k,\kp}$,
we can obtain the dynamic shear moduli with the help of (\ref{koreGpGPP})
where the summation is taken over $2\le k,\kp\le s_m$.
Note that the number of unknowns in (\ref{kretoek2}) is $s_m(s_m-1)$;
when $s_m=4$, for example, there are 12 unknowns:
$g^{\prime}_{2,2}, g^{\prime}_{3,2}, g^{\prime}_{3,3}, g^{\prime}_{4,2}, 
g^{\prime}_{4,3}, g^{\prime}_{4,4}$
and 
$g^{\prime\prime}_{2,2}, g^{\prime\prime}_{3,2}, 
g^{\prime\prime}_{3,3}, g^{\prime\prime}_{4,2}, g^{\prime\prime}_{4,3}, 
g^{\prime\prime}_{4,4}$.


Fig.\ref{getatau1} (i) shows the dynamic shear moduli reduced by $nk_BT$
as a function of the unitless frequency.
The reduced polymer concentration is changed from curve to curve
for a maximum multiplicity fixed at $s_m=10$.
It appears that they are Maxwellian with a single relaxation time.
To see the detail,
the plateau modulus
$G_{\infty}=\lim_{\omega\to\infty}G^{\prime}(\omega)$
reduced by $nk_BT$ is drawn in Fig.\ref{getatau1} (ii) as a function of $c$.
The reduced plateau modulus
increases 
as $c^2$ 
irrespective of the value of $s_m$
when $c$ is small\footnote{When $c$ is small, 
the extent of reaction is proportional to $c$, i.e.,
$\alpha\sim c/(1+c)\sim c$. 
Therefore, we find $G_{\infty}/nk_BT=\alpha^2\sim c^2$.}
and approaches to 1 with increasing $c$.
%
%
Fig.\ref{getatau1} (iv) shows the relaxation time
determined from the peak position of $G^{\prime\prime}$
as a function of $c$.
The relaxation time 
becomes close to $1/(4\beta)$ 
in the limit of low reduced concentration
because most active chains belong to junctions with $k=2$
in this limit. 
%
%
\begin{figure}[t]
\begin{center}
\includegraphics*[scale=0.6]{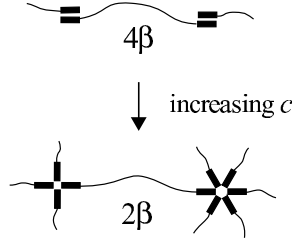}
\end{center}
\vspace*{-0.5cm}
\caption{
The breakage rate of the chain 
linking 
two $k=2$ junctions (upper figure)
is $4\beta$ (see Fig.\ref{pair}).
On the other hand, the breakage rate of the chain 
connecting 
two junctions
with the multiplicity $k\ge3$ (lower figure)
is $2\beta$ because
the chain 
debridges
if its one of two ends are dissociated from the junction.
}
\label{pair2}
\end{figure}
On the other hand, $\tau$ approaches to $1/(2\beta)$ with increasing $c$
because more active chains tend to connect with junctions with $k\ge 3$
when $c$ is large (see Fig.\ref{pair2}).
Thus $\tau$ increases with increasing $c$.
The zero-shear viscosity
$\eta_0=\lim_{\omega\to0}G^{\prime\prime}(\omega)/\omega$
reduced by $nk_BT/\beta$ is shown in Fig.\ref{getatau1} (iii)
as a function of $c$.
The zero-shear viscosity is roughly estimated to be $\eta_0\sim G_{\infty}\tau$
(or $\eta_0/(nk_BT/\beta) \sim G_{\infty}/(nk_BT)\beta\tau$),
and hence
the reduced viscosity starts to increase at $c=0$
and approaches to 0.5 with increasing $c$.


\begin{figure}[t]
\begin{center}
\includegraphics*[scale=0.5]{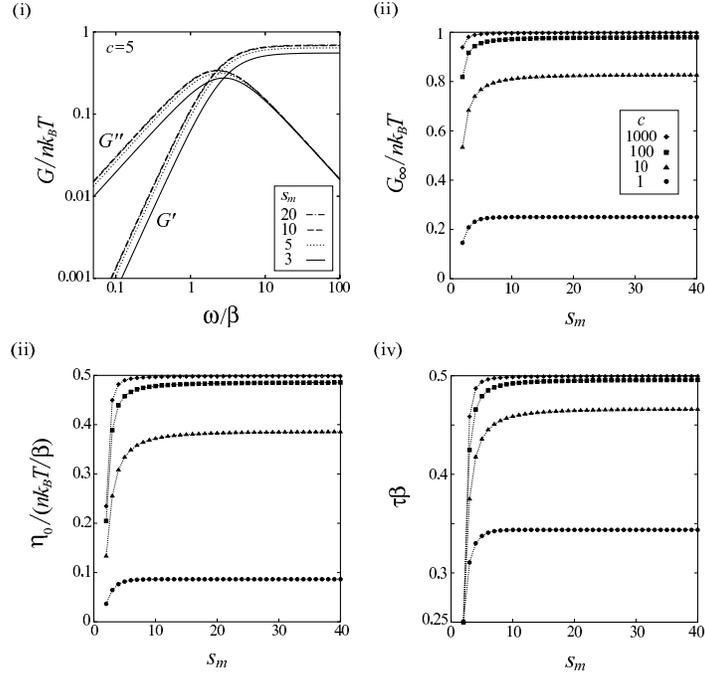}
\end{center}
\vspace*{-0.5cm}
\caption{
(i) The dynamic shear moduli (reduced by $nk_BT$) 
for the saturating junction model
as a function of the frequency.
The maximum multiplicity $s_m$ is varying from curve to curve 
for a reduced concentration fixed at $c=5$.
(ii) The reduced plateau modulus, 
(iii) reduced zero-shear viscosity, 
and (iv) relaxation time
plotted against the maximum multiplicity
for several reduced concentrations.
}
\label{g3}
\end{figure}

Fig.\ref{g3} (i) shows the 
reduced shear moduli
for several maximum multiplicity $s_m$.
The reduced plateau modulus, reduced zero-shear viscosity
and relaxation time
are also plotted in Fig.\ref{g3} as a function of $s_m$.
Let us comment on 
the relaxation time here. 
The relaxation time is determined from the ratio between the number of 
active chains
incorporated into the junction with multiplicity $k\ge 3$ and that with $k= 2$.
When $s_m=2$, for example, all active chains 
belong to $2$-junctions, and therefore we have $\tau=1/(4\beta)$ 
irrespective of the value of $c$.
When $s_m\ge 3$, 
more chains can connect to junctions with $k\ge 3$, so that 
the fraction $q_2$ of chains belonging to $2$-junctions decreases
(this tendency can also be confirmed from Fig.\ref{chi2}).
Thus $\tau$ increases with increasing $s_m$ and
approaches to a fixed value for each $c$.
As for 
the dependence of 
$G_{\infty}/nk_BT=\alpha^2$
on $s_m$, 
see \ref{satuequ} for reference.

\subsection{Fixed Multiplicity Model}

\begin{figure}[t]
\begin{center}
\includegraphics*[scale=0.5]{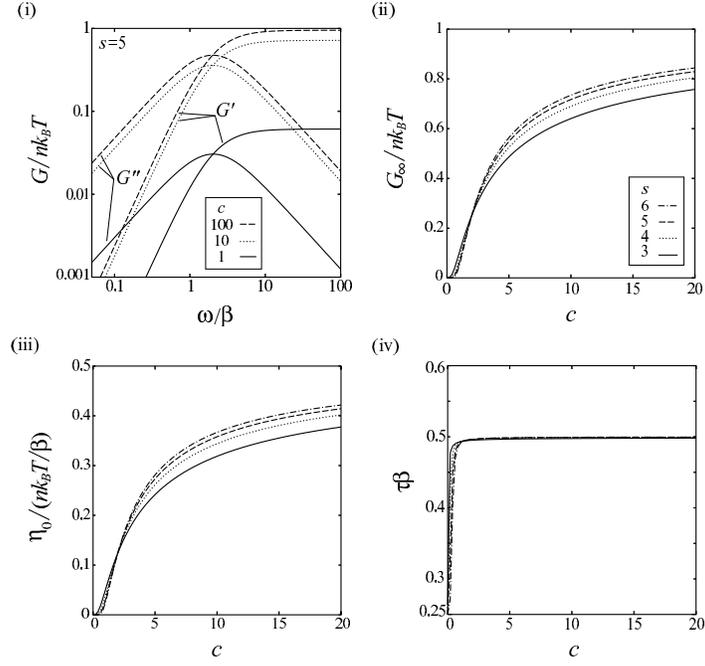}
\end{center}
\vspace*{-0.5cm}
\caption{
(i) The dynamic shear moduli (reduced by $nk_BT$)
for the fixed multiplicity model
as a function of the frequency.
The reduced concentration is varying from curve to curve 
while the multiplicity is fixed at $s=5$.
(ii) the reduced plateau modulus, 
(iii) reduced zero-shear viscosity, 
and (iv) relaxation time
plotted against the reduced concentration.
}
\label{g5}
\end{figure}

Next, we consider the fixed multiplicity model, i.e., 
the multiplicity can take only $k=1$ (unreacted) and $k=s$ (reacted)
for all junctions.
This condition can be approximately attained by employing (\ref{deltasyou})
for $P_k$, i.e.,
\bea
& &P_k= \beta z k\left\{
\begin{array}{@{\,}ll}
\delta          & (1\le k< s-1) \\
\delta^{-(s-2)} & (k=s-1) \\
0 & (k> s-1)
\end{array}
\right.,
\label{newPk3}
\eea
where we are setting $\delta=0.01$. 
By putting (\ref{newPk3}) into (\ref{kretoek2}) 
and by solving a simultaneous equation,
we can obtain the dynamic shear moduli for the (approximate) fixed multiplicity model.

Fig.\ref{g5} shows the dependence of 
the reduced shear moduli on $c$. 
The reduced plateau modulus (figure (ii))
increases with increasing $s$ for $c>2$, but it decreases for $c<2$
according to the $s$ dependence of $\alpha$ (see Fig.\ref{alpha4}).
The relaxation time (figure (iv))
is almost $1/(2\beta)$ irrespective of the values of $s(\ge 3)$ and $c$
because almost all junctions take the multiplicity larger than or equal to three.
However, there is a small number of 2-junctions
in the system, especially when $c$ is close to 0, 
due to the finite $\delta$.
Thus, the relaxation time becomes smaller than $1/(2\beta)$
(but larger than $1/(4\beta)$) for $c$ close to 0.
The reduce zero-shear viscosity (figure (iii)) is approximately half the 
reduced plateau modulus because $\beta\tau$ is about 0.5.


\section{Summary and Discussions}
\label{secvidis}

In this paper, we developed a theory for transient networks
with multiple junctions.
One of our objectives is to investigate relations between 
rheological (and equilibrium) properties of the network
and the junction multiplicity 
in the absence of looped chains.
We assumed that the connection rate of a functional group 
is proportional to the 
volume fraction of junctions to which
it is going to connect,
and showed that the law of mass action holds. 
%
As an first attempt, we defined active chains locally, i.e.,
chains whose both ends are connected with any other 
functional groups are elastically effective. 
%
The dynamic shear moduli are well described in terms of the Maxwell model
characterized by a single relaxation time and the high-frequency plateau modulus
(and the zero-shear viscosity).
%
The reduced moduli depend on thermodynamic quantities such as 
polymer concentration and temperature
through the reduced polymer concentration $c$.
The plateau modulus and the zero-shear viscosity 
increase 
nonlinearly 
at small $c$
and proportional to $c$
when $c$ is large.
The relaxation time also increases with increasing $c$ due to the presence of 
pairwise junctions at small $c$.

The modulus, viscosity and relaxation time start to rise at $c=0$. 
This indicates that the critical concentration for the sol/gel transition 
is $c^*=0$ in this model.
In other words, the system is always in the postgel regime
irrespective of the amount of polymers.
This unfavorable result 
can be ascribed to
the lack of global information 
in the current definition of
active chains.
%


The present theory reduces to the TE theory 
in the high reduced concentration limit $c\to\infty$.
Let us take a summation
over 
$2\le k,\kp\le s_m$
in (\ref{activeeq}).
Then we have
\bea
& &\frac{\partial F(\br,t)}{\partial t}+\nabla\cdot\left(\dot{\br}F(\br,t)\right)
=-2\beta(r)F(\br,t)-B_2(t)\sum_{k= 2}^{s_m}\Bigl(F_{k,2}(\br,t)+F_{2,k}(\br,t)\Bigr)
\non\\& &\hspace*{4.95cm}
+\Bigl(\sum_{k= 1}^{s_m-1}p_{k}(t)+p_1(t)\Bigr)
\nu^d(t)f_0(\br), \label{wangsiki}
\eea
where
\bea
F(\br,t)\equiv\sum_{k= 2}^{s_m}\sum_{\kp= 2}^{s_m}F_{k,\kp}(\br,t)
\eea
is the total number of active chains with the head-to-vector $\br$,
and
\bea
\nu^d(t)\equiv \sum_{k= 2}^{s_m}\bigl(\nu_{k,1}(t)+\nu_{1,k}(t)\bigr)
\eea
is the total number of dangling chains.
(We are assuming that the dissociation rate does not depend on the junction 
multiplicity as in the text.)
The second term of the right-hand side in (\ref{wangsiki})
and $p_1$ inside the second parentheses 
are related with the annihilation/creation process of 2-junctions.
As shown below, 
the relative amount of 2-junctions becomes negligible 
in the high $c$ limit, and hence
effects of these terms disappear.
%
%
Let us consider the case that 
a small shear deformation is applied to the system as discussed in the text.
Upon integration with respect to $\br$, eq.(\ref{wangsiki}) becomes
\bea
0=-2\beta(\nu^{\eff}+\sum_{k\ge 2}\nu_{k,2})
+\Bigl(\sum_{k\ge 1}p_{k}+p_1\Bigr).
\label{sekibunat}
\eea
The terms $\sum_{k\ge 2}\nu_{k,2}$ and $p_1$ associated with 2-junctions
satisfy the following relation:
\bea
& &\frac{\sum_{k\ge 2}\nu_{k,2}}{\nu^{\eff}}=
\frac{p_1}{\sum_{k\ge 1}p_{k}}=\frac{q_2}{\alpha}.
\label{qs2as}
\eea
\begin{figure}[t]
\begin{center}
\includegraphics*[scale=0.5]{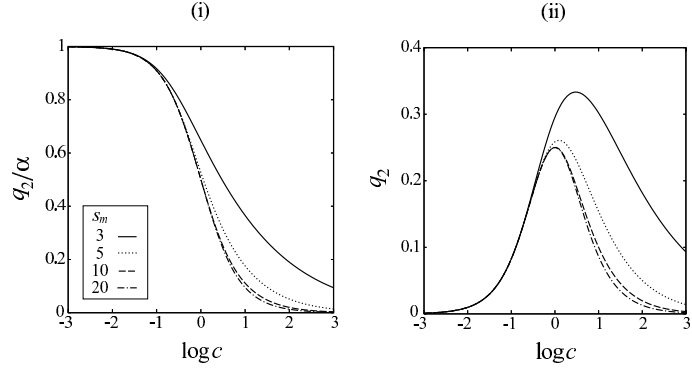}
\end{center}
\vspace*{-0.5cm}
\caption{
(i) Effects of the terms in (\ref{wangsiki}) and (\ref{sekibunat})
associated with the creation/annihilation process of 2-junctions,
and
(ii) the fraction of chains belonging to 2-junctions
plotted against the reduced polymer concentration.
The maximum multiplicity is varying from curve to curve in both figures.
}
\label{q2}
\end{figure}
With increasing the reduced concentration,
$q_2/\alpha$ approaches to zero 
as shown in Fig.\ref{q2} (i),\footnote{Eq. (\ref{qs2as}) 
can be approximately expressed as $1/(1+c)$ when $s_m$ is large.}
and hence the right-hand side of (\ref{sekibunat}) becomes close to
$-2\beta\nu^{\eff}+\sum_{k\ge 1}p_{k}$.
%
It indicates that (\ref{wangsiki}) reduces to
\bea
& &\frac{\partial F(\br,t)}{\partial t}+\nabla\cdot\left(\dot{\br}F(\br,t)\right)
=-2\beta F(\br,t)
+p\nu^df_0(\br) \label{wangsiki2}
\eea
in the high $c$ limit
where 
\bea
p\equiv \sum_{k\ge 1}p_{k}
\label{ptteigi}
\eea
is the probability that an unreacted group connects to {\it any} junctions
per unit time.
Eq. (\ref{wangsiki2}) is 
equivalent to the fundamental equation of the TE theory
when we regard $p$ as a constant and assume that isolated chains are 
absent.\footnote{$2\beta$ 
is the transition rate from active chains
to dangling chains. This quantity is denoted as $\beta$ 
in the TE theory \cite{tanaed1,tanaed2}.}

\begin{figure}[t]
\begin{center}
\includegraphics*[scale=0.5]{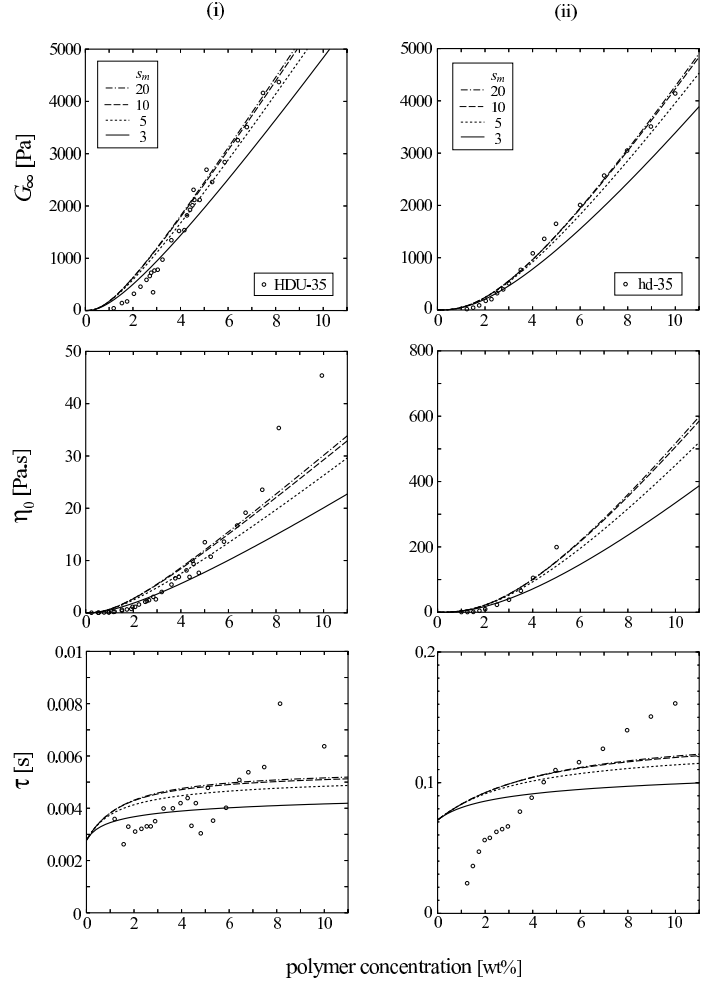}
\end{center}
\vspace*{-0.5cm}
\caption{
Comparison between theoretical results for the saturating junction model
and experimental data obtained for
(i) telechelic PEO with narrow molecular weight distributions 
($Mw\!=\!35$kg/mol) fully end-capped with C$_{16}$ alkanes
reported by Pham {\it et al.} \cite{russel} (called HDU-35)
and
(ii) hydrophobically modified ethylene oxide-urethane copolymers (HEUR)
with similar molecular weight end-capped with the same hydrophobes
reported by Annable {\it et al.} \cite{annable1} (called hd-35 after ref.\cite{meng}).
Theoretical curves for the zero-shear viscosity and the relaxation time are drawn with
$\beta=90$ (1/sec) for HDU-35 and $\beta=3.5$ (1/sec) for hd-35.\protect\footnote{A 
discrepancy in the value of $\beta$ for the same alkanes 
might stem from the difference
in the coupling agents between the alkanes and the PEO backbone \cite{russel,meng}.}
A factor $\xi$ (see text)
is assumed to be 1 for HDU-35 and 0.35 for hd-35, respectively.
}
\label{exhikaku}
\end{figure}

In Fig.\ref{exhikaku},
theoretically obtained plateau modulus, zero-shear viscosity and relaxation time
for the saturating junction model
are compared with
with experimental data 
for aqueous solutions of telechelic PEO
end-capped with C$_{16}$ alkanes \cite{annable1,russel}.
The reduced concentration $c$ 
used in the theory was converted
into the polymer concentration in weight percentage $c_w$ 
through a relation $c=\xi c_w$, where $\xi\equiv (2000N_A/M)\lambda v_0$
($N_A$ is Avogadro's number).
We see that both agree fairy well with each other 
for larger $s_m$ except for the relaxation time of HEUR aqueous solutions.
According to Annable {\it et at.} \cite{annable1},
a concentration dependence of the relaxation time
is strongly affected by 
the amount of {\it superbridges} 
formed by connecting several primary bridges in series 
because they have shorter lifetime due to a number of possible 
disengagement points inside.
Actually, we can infer from Fig.\ref{q2} (i) that
a considerable fraction of active chains are forming superbridges
at low concentration;
a fraction of active chains $(q_2/\alpha)^2$
whose both ends are connected with 2-junctions
becomes close 1 when $c\to0$
although both $\alpha$ and $q_2$ approach to 0 
(see Fig.\ref{alpha2} (ii) and Fig.\ref{q2} (ii) respectively).
In order to treat such superbridges in more appropriate manner,
we need to consider a global structure of the network
by introducing a concept of the path connectivity to the 
infinite network according to refs.\cite{tanaishi,graessley}.
The effect of superbridges on the viscoelasticities will be discussed in detail
in the following paper.


\appendix


\section{Derivation of (\ref{dan2no2})}
\label{appen1}

\begin{figure}[t]
\begin{center}
\includegraphics*[scale=0.6]{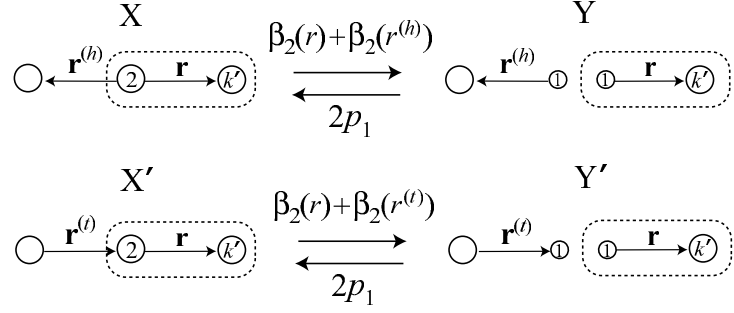}
\end{center}
\vspace*{-0.5cm}
\caption{
Transitions between the $k=1$ chain and the $k=2$ chain.
}
\label{hitots5}
\end{figure}

The number of the state X depicted in Fig.\ref{hitots5} 
(corresponding to the state A in Fig.\ref{hitotsu} with $k_h=1$ and $k_t=0$)
is
\bea
{\cal F}_{2,\kp}(\br,\br^{(h)};t)
=\frac{1}{2}F_{2,\kp}(\br,t)f_2^{(h)}(\br^{(h)},t).
\eea
The $(1,\kp)$-chain is generated from the state X
if
the head of the $(2,\kp)$-chain 
{\it or} the head of another chain
is detached
from the $2$-junction 
(see the transition from the state X to state Y in Fig.\ref{hitots5}).
Then the increment of $(1,\kp)$-chains due to this process is estimated to be
\bea
& &\int \!d\br\!\int\! d\br^{(h)}\Bigl(\beta_2(r)+\beta_2(r^{(h)})\Bigr)
{\cal F}_{2,\kp}(\br,\br^{(h)};t) \non\\
& &=\frac{1}{2}\int \!d\br~\beta_2(r)F_{2,\kp}(\br,t)
+\frac{1}{2}\left( \int \!d\br~\beta_2(r)f_2^{(h)}(\br,t) \right)\nu_{2,\kp}(t).
\label{dank1nosiki}
\eea
Similarly,
the increment of 
$(1,\kp)$-chains
as a result of the transition from the state
X$^\prime$ (in Fig.\ref{hitots5}) to Y$^\prime$ is 
written as
\bea
\frac{1}{2}\int \!d\br~\beta_2(r)F_{2,\kp}(\br,t)
+\frac{1}{2}\left( \int \!d\br~\beta_2(r)f_2^{(t)}(\br,t) \right)\nu_{2,\kp}(t).
\label{dank1nosiki2}
\eea
By adding (\ref{dank1nosiki}) and (\ref{dank1nosiki2}),
we obtain (\ref{dan2no2}).





\bibliographystyle{aipprocl}

\begin{thebibliography}{999}


%
%


\bibitem{winyek}
M.A.Winnik, A.Yekta,
Associative polymers in aqueous solution,
Curr. Opin. Collod Interface Sci.,
2 (1997) 424-436.



\bibitem{jen1}
R.D.Jenkins, C.A.Silebi,M.S.El-Aasser,
Steady-shear and linear viscoelastic material properties of 
model associative polymer solutions,
ACS Symp. Ser.,
462 (1991) 222-233.


\bibitem{annable1}
T.Annable, R.Buscall, R.Ettelaie, D.Whittlestone,
The rheology of solutions of associating polymers: 
Comparison of experimental behavior with transient network theory,
J. Rheol.,
37 (1993) 695-726.


\bibitem{jen2}
R.D.Jenkins, D.R.Bassett, C.A.Silebi, M.S.El-Aasser,
Synthesis and characterization of model associating polymers,
J. Appl. Polym. Sci.,
58 (1995) 209-230.


\bibitem{winnik}
A.Yekta, B.Xu, J.Duhamel, H.Adiwidjaja, M.A.Winnik,
Fluorescence studies of associating polymers in water:
determination of the chain end aggregation number and 
a model for the association process,
Macromolecules,
28 (1995) 956-966.





%
%
%


\bibitem{tam}
K.C.Tam, R.D.Jenkins, M.A.Winnik, D.R.Bassett,
A structural model of hydrophobically modified urethane-ethoxylate (HEUR)
associative polymers in shear flows,
Macromolecules,
31 (1998) 4149-4159.



\bibitem{winnik1}
O.Vorobyova, A.Yekta, M.A.Winnik, W.Lau,
Fluorescent probe studies of the association in an 
aqueous solution of a hydrophobically modified poly(ethylene oxide),
Macromolecules,
31 (1998) 8998-9007.


\bibitem{russel0}
Q.T.Pham, W.B.Russel, J.C.Thibeault, W.Lau,
Micellar solutions of associative triblock copolymers: 
Entropic attraction and gas-liquid transition,
Macromolecules,
32 (1999) 2996-3005.


\bibitem{russel}
Q.T.Pham, W.B.Russel, J.C.Thibeault, W.Lau,
Micellar solutions of associative triblock copolymers: 
The relationship between structure and rheology,
Macromolecules,
32 (1999) 5139-5146.


\bibitem{ng}
W.K.Ng, K.C.Tam, R.D.Jenkins,
Lifetime and network relaxation time of a HEUR-C20 associative polymer system,
J. Rheol.
44 (2000) 137-147.


\bibitem{serero0}
Y.S\'{e}r\'{e}ro, V.Jacobsen, J.-F.Berret, R.May,
Evidence of non-linear chain stretching in the rheology of transient networks,
Macromolecules 
33 (2000) 1841-1847.


\bibitem{ma}
S.X.Ma, S.L.Cooper,
Shear thickening in aqueous solutions of hydrocarbon end-capped poly(ethylene oxide),
Macromolecules,
34 (2001) 3294-3301.

\bibitem{ma2}
S.X.Ma, S.L.Cooper,
Effect of polydispersity on viscoelasticity and shear thickening in 
aqueous solutions of hydrocarbon end-capped poly(ethylene oxide),
Macromolecules,
35 (2002) 2024-2029.


\bibitem{serero}
D.Calvet, A.Collet, M.Viguier, J.-F.Berret, Y.S\'{e}r\'{e}ro,
Perfluoroalkyl end-capped poly(ethylene oxide). 
Synthesis, characterization, and rheological behavior in aqueous solution,
Macromolecules,
36 (2003) 449-457.


\bibitem{nicolai}
F.Lafleche, D.Durand, T.Nicolai
Association of adhesive spheres formed by hydrophobically end-capped PEO.
1. Influence of the presence of single end-capped PEO,
Macromolecules,
36 (2003) 1331-1340.


\bibitem{mewis1}
L.Pellens, R.G.Corrales, J.Mewis, 
General nonlinear rheological behavior of associative polymers,
J. Rheol.,
48 (2004) 379-393.


\bibitem{mewis2}
L.Pellens, K.H.Ahn, S.J.Lee, J.Mewis,
Evaluation of a traneient network model for telechelic associative polymers,
J. Non-Newtonian Fluid Mech.,
121 (2004) 87-100.



\bibitem{kujawa}
P.Kujawa, H.Watanabe, F.Tanaka, F.M.Winnik,
Amphiphilic telechelic poly(N-isopropylacrylamide) in water: from micelles to gels,
Eur. Phys. J. E,
17 (2005) 129-137.







\bibitem{semenov}    
A.N.Semenov, J.-F.Joanny, A.R.Khokhlov,
Associating polymers: Equilibrium and linear viscoelasticity,
Macromolecules,
28 (1995) 1066-1075.



\bibitem{tanaed1}
F.Tanaka, S.F.Edwards,
Viscoelastic properties of physically crosslinked networks.
Transient network theory,
Macromolecules,
25 (1992) 1516-1523.


\bibitem{tanaed2}
F.Tanaka, S.F.Edwards,
Viscoelastic properties of physically crosslinked networks,
Parts 1-3,
J. Non-Newtonian Fluid Mech.,
43 (1992) 247-271, 273-288, 289-309.


\bibitem{wang}
S.Q.Wang,
Transient network theory for shear-thickening fluids and 
physically cross-linked systems,
Macromolecules,
25 (1992) 7003-7010.


\bibitem{mar0}
G.Marrucci, S.Bhargava, S.L.Cooper,
Models of shear-thickening behavior in physically cross-linked networks,
Macromolecules,
26 (1993) 6483-6488.


\bibitem{mar1}
A.Vaccarro, G.Marrucci,
A model for the nonlinear rheology of associating polymers,
J. Non-Newtonian Fluid Mech.,
92 (2000) 261-273.


\bibitem{meng}
X.-X.Meng, W.B.Russel,
Rheology of telechelic associative polymers in aqueous solutions
J. Rheol.,
50 (2006) 189-205.


\bibitem{innon}
T.Indei, 
Necessary conditions for shear thickening
in associating polymer networks,
J. Non-Newtonian Fluid Mech.,
to appear (2006).


\bibitem{gt}
M.S.Green, A.V.Tobolsky,
A new approach to the theory of relaxing polymeric media,
J. Chem. Phys.,
14 (1946) 80-92.


\bibitem{intana1}
T.Indei, F.Tanaka,
Rheological study of transient polymer networks crosslinked by 
two-component associative groups, - inversion of the gel skeletal structure,
J. Rheol., 
48 (2004) 641-661.

\bibitem{intana2}
T.Indei, F.Tanaka,
Theory of transient polymer networks crosslinked by two-component associative groups,
Nihon Reoroji Gakkaishi (J Soc Rheol, Jpn), 
32, (2004) 285-293.




\bibitem{tanastock}
F.Tanaka, W.H.Stockmayer,
Thermoreversible gelation with junctions of variable multiplicity,
Macromolecules,
27 (1994) 3943-3954.



\bibitem{fransur}
W.Binana-Limbele, F.Clouet, J.Franc\c{o}is,
Hydrophobically end-capped poly(ethylene oxide) urethanes.
3. Effect of sodium dodecyl-sulfate on their association in aqueous-solution,
Colloid Polym. Sci.,
271 (1993) 748-758.


\bibitem{annable2}
T.Annable, R.Buscall, R.Ettelaie, P.Shepherd, D.Whittlestone,
Influence of surfactants on the rheology of associating polymers in solution,
Langmuir,
10 (1994) 1060-1070.


\bibitem{winniksur}
K.Zhang, B.Xu, M.A.Winnik, P.M.Macdonald,
Surfactant interactions with HEUR associating polymers 
J. Phys. Chem.,
100 (1996) 9834-9841.




















\bibitem{tanaishi}
F.Tanaka, M.Ishida,
Elastically effective chains in transient gels with multiple junctions,
Macromolecules,
29 (1996) 7571-7580.



\bibitem{graessley}
D.S.Pearson, W.W.Graessley,
The structure of rubber networks with multifunctional junctions,
Macromolecules,
11 (1978) 528-533.

















\end{thebibliography}

\end{document}